\documentclass{aastex631}
\let\tablenum\relax 
\usepackage[print-unity-mantissa=false]{siunitx}
\usepackage{amsmath}
\usepackage{bm}
\usepackage{commath}
\usepackage[safe]{tipa}
\usepackage{multirow}
\usepackage{ulem}

\DeclareSIUnit\parsec{pc} 
\DeclareSIUnit\year{yr}
\DeclareSIUnit\arcsecond{as} 

\DeclareRobustCommand{\okina}{%
  \raisebox{\dimexpr\fontcharht\font`A-\height}{%
    \scalebox{0.8}{`}%
  }%
}
\newcommand{\Ou}{{\okina}Oumuamua}
\newcommand{\FeH}{[\text{Fe}/\text{H}]}
\newcommand{\MH}{[\text{M}/\text{H}]}
\newcommand{\fHHO}{f_{\mathrm{H}_{2} \mathrm{O}}}
\newcommand{\figwidth}{0.49}

\received{2024 Feb 07}
\revised{2024 November 29}
\accepted{2024 December 04}

\shorttitle{ISO Chemodynamics with Gaia}
\shortauthors{Hopkins et al.}

\begin{document}

\title{Predicting Interstellar Object Chemodynamics with Gaia}

\correspondingauthor{Matthew Hopkins}
\email{matthew.hopkins@physics.ox.ac.uk}

\author[0000-0001-6314-873X]{Matthew J. Hopkins}
\affiliation{Department of Physics, University of Oxford, Denys Wilkinson Building, Keble Road, Oxford, OX1 3RH, UK}
\affiliation{School of Physical and Chemical Sciences|Te Kura Mat\={u}, University of Canterbury,
Private Bag 4800, Christchurch 8140,
New Zealand}

\author[0000-0003-3257-4490]{Michele T. Bannister}
\affiliation{School of Physical and Chemical Sciences|Te Kura Mat\={u}, University of Canterbury,
Private Bag 4800, Christchurch 8140,
New Zealand}

\author[0000-0001-5578-359X]{Chris Lintott}
\affiliation{Department of Physics, University of Oxford, Denys Wilkinson Building, Keble Road, Oxford, OX1 3RH, UK}

\begin{abstract}

The interstellar object population of the Milky Way is a product of its stars.
However, what is in fact a complex structure in the Solar neighbourhood has traditionally in ISO studies been described as smoothly distributed.
Using a debiased stellar population derived from the \textit{Gaia} DR3 stellar sample, we predict that the velocity distribution of ISOs is far more textured than a smooth Gaussian. 
The moving groups caused by Galactic resonances dominate the distribution.
1I/\Ou\ and 2I/Borisov have entirely normal places within these distributions; 1I is within the non-coeval moving group that includes the Matariki (Pleiades) cluster, and 2I within the Coma Berenices moving group.
We show that for the composition of planetesimals formed beyond the ice line, these velocity structures also have a chemodynamic component.
This variation will be visible on the sky.
We predict that this richly textured distribution will be differentiable from smooth Gaussians in samples that are within the expected discovery capacity of the Vera C. Rubin Observatory.
Solar neighbourhood ISOs will be of all ages and come from a dynamic mix of many different populations of stars, reflecting their origins from all around the Galactic disk.
 
\end{abstract}


\section{Introduction} 
\label{sec:intro}

The three-dimensional velocity distribution of interstellar objects (ISOs) is both complex, and a valuable parameter space with which to understand the Galactic small-body population.
As it is a product of their kinematics, it correlates with the chemistry of their progenitor stars, their ejection mechanisms, and the effects of Galactic dynamical processes including encounters with spiral arms and giant molecular clouds.
    
In the local environment --- the Solar neighbourhood, and specifically the region which our observable volume of the Solar System samples --- the velocity distribution affects a range of factors.  
Locally, it can affect ISO detectability within the Solar System volume \citep{Francis_2005}. 
As a consequence it affects estimates of the total Galactic population size, therefore affecting inference on the various ISO production mechanisms \citep{Moro-Martin_2009}.
The breadth of ISO velocities also limits the potential for encounter and sampling visits to ISOs by spacecraft, which have finite $\Delta v$ \citep{Moore_2021,Jones_2024}.
Finally, it also affects what can be inferred of the origin history of the two currently detected ISOs, 1I/\Ou\ and 2I/Borisov.

Previous works considering a velocity distribution for ISOs consistently assume that it is equal to the stellar distribution, and assume that the stellar distribution is a multivariate Gaussian \citep{Whipple_1975,Sekanina_1976,McGlynn_1989,Stern_1990,Cook_2016,Engelhardt_2017,Seligman_2018,Forbes_2019,Marceta_2020,Hoover_2022,Marceta_2023b,Marceta_2023a}.
However, \citet{Binney_1998} caution that the local stellar velocity distribution is not well described by a Gaussian; in particular, being smooth, these distributions will miss any structure and features of the velocity distribution of objects in the Solar neighbourhood.

The stellar velocity distribution is not smooth: it has strong overdensities called moving groups and branches, known since \cite{Eddington_1906} and well-studied since \textit{Hipparcos} \citep{Dehnen_1998,Skuljan_1999}.
These are not coeval \citep{Nordstrom_2004,Bensby_2007,Famaey_2008,Antoja_2008}, so while originally thought to be dispersed star clusters, they are now understood to be temporary structures, sculpted by resonances with the Galaxy's spiral arms and bar --- which themselves may be transient \citep{ Quillen_2011, Ramos_2018, Antoja_2018,Michtchenko_2018,Hunt_2018, Hunt_2019, Trick_2021, Lucchini_2023}.
In the age of \textit{Gaia}, the local stellar velocity distribution is now known with exquisite precision.
Additionally, protoplanetary disk modelling is now able to link stellar chemcial abundances to aspects of planetesimal composition, for example \cite{Santos_2015} who predict the iron mass fraction of exoplanets and \cite{Bitsch_2020} who predict planetesimal composition both inside and outside the water ice line.
Thus variations in the physical composition of ISOs, such as their water mass fraction, can be mapped to their home system's stellar metallicity --- making it possible to use ISOs as sensitive tracers of the Galactic star formation history \citep{Lintott_2022}.
This strongly suggests that combining predictions of ISO physical compositions with their three-dimensional velocity distributions will reveal structure. For consistency with Galactic evolution terminology, we refer to this as their `chemodynamics' \citep[as used in e.g.][]{Burkert_1987,Binney_2024}.

We use current knowledge of the local stellar population within 200 pc of the Sun to model the local ISO distribution.
Using the inference framework developed in \citet{Hopkins_2023} and the \textit{Gaia} DR3 measurements, we explore the kinematics, specifically the velocity distributions, and the chemodynamics of the local population structure of interstellar objects.
We term this model the \=Otautahi-Oxford model.
This is the first such study to account for metallicity effects and stellar death, or examine the 3D kinematic structures, though \cite{Eubanks_2021} estimate the ISO speed distribution from \textit{Gaia} eDR3 by applying a volume sampling rate to a subset of the \textit{Gaia} Catalogue of Nearby Stars \citep{GaiaCollaboration_2021} with radial velocities.

We show that the local velocity distribution of ISOs is richly featured and the velocity of an ISO correlates with the metallicity and age of its origin star, and therefore its own composition and age.
We demonstrate the sample sizes needed to distinguish between our predicted ISO velocity distribution and smooth Gaussian distributions.
This chemodynamic texture will be perceptible by upcoming Solar System surveys, such as the Vera C. Rubin Observatory Legacy Survey of Space and Time \citep[LSST;][]{Ivezic_2019}.

\section{Method}
\label{sec:method}

In \citet{Hopkins_2023} we used positions and elemental abundances of stars from APOGEE \citep{Jonsson_2020} for modelling the ISO spatial distribution over a large swath of the Galactic disk; here our focus is the velocity distribution in the Solar neighbourhood, so we use \textit{Gaia} \citep{GaiaCollaboration_2016}.
We first use \textit{Gaia} measurements of stellar proper motion, radial velocity, metallicity and age to calculate the distribution in these properties of the stellar population within \qty{200}{\parsec} of the Sun.
We account for the survey's selection effects on each of these measurements.
First, we use the methods of \cite{Cantat-Gaudin_2023} and \cite{Castro-Ginard_2023} to calculate a selection function in color and magnitude; we then calculate our own effective selection function in age, metallicity and distance, simultaneously accounting for stellar death (\S~\ref{sec:SF}). 
As in \citet{Hopkins_2023}, this lets us recover the distribution of the \textit{sine morte}\footnote{We continue use of this from \citet{Hopkins_2023}: \textit{sine morte}, Latin for ``without death'', [\textprimstress si\textlengthmark ne \textprimstress m\textopeno rte] IPA pronunciation, or `seen-ay mort-ay'.} stellar population: what the stellar population would be if stars did not die (\S~\ref{sec:sine_morte}).
From our debiased stellar distribution, we then predict the ISO distribution: we combine this stellar population with the protoplanetary disk chemical model of \cite{Bitsch_2020}, and the assumptions in \S~\ref{sec:stars_to_isos} on the ejection of planetesimals and their motion in the Galaxy.
This gives us a predicted distribution of ISOs in the Solar neighbourhood in velocity, age and composition.
Finally, to predict the distribution of ISOs entering the inner Solar system, where they will be detectable by sky surveys such as the Vera C. Rubin Observatory's LSST, we apply a gravitationally-focussed volume sampling rate to account for the motion of ISOs relative to the Sun (\S~\ref{sec:grav_focussing}).

\subsection{Gaia and its Selection Function}
\label{sec:SF}

\textit{Gaia} DR3 provides a comprehensive dataset of the stars in our part of the Galaxy \citep{GaiaCollaboration_2023b}.
We first obtain a subset of \textit{Gaia} DR3, requiring the presence of high-precision measurements of velocity, metallicity, and age in our sample.
We define the Solar neighbourhood as a sphere around the Sun out to the somewhat arbitrary distance of \qty{200}{\parsec}, similar in scale to the distances used by \cite{Antoja_2018} and \cite{Recio-Blanco_2023}, so we take stars with \textit{Gaia} trigonometric parallax \(\geq\qty{5}{\milli\arcsecond}\).\footnote{N.B. Due to a parallax bias of size \(\sim \qty{10}{\micro\arcsecond}\) \citep{Lindegren_2021} this will not correspond to exactly \qty{200}{\parsec}.}
To ensure these parallaxes are accurate, we require \(\mathtt{parallax\_over\_error}>10\), and in order to calculate the 3D velocities of each star, we require a radial velocity to have been measured and have an error of less than \qty{5}{\kilo\meter\per\second}.
We require the availability of GSP-Spec\footnote{General Stellar Parametrizer from Spectroscopy} metallicity \(\MH\) and GSP-Spec-based FLAME\footnote{Final Luminosity Age Mass Estimator} ages, both of which are measured using data from the Radial Velocity Spectrometer \citep[RVS;][]{Cropper_2018}: \(\MH\) is measured by fitting RVS spectra to synthetic spectra from a five-parameter stellar atmosphere model \citep{Recio-Blanco_2023}, and GSP-Spec-based FLAME age is calculated from these atmospheric parameters and isochrone fitting \citep{Fouesneau_2023}.
Comparing to literature values, GSP-Spec \(\MH\) estimates show no bias and a dispersion of 0.13 dex \citep{Recio-Blanco_2023}, while FLAME ages have mean biases and dispersions of only 0.1 to 0.3 Gyr and 0.25 Gyr respectively \citep{Fouesneau_2023}.
To ensure accurate ages, we remove giants from the sample by requiring \(\mathtt{flags\_flame\_spec}=0\) \citep{CreeveyLebreton_2022,Creevey_2023,Fouesneau_2023}.
These measurements have the highest accuracy available; while only present for bright stars (\(G\lesssim14\)), within \qty{200}{\parsec} we can account for the selection effects these cuts induce.
This yields a sample of \num{201863} observed living stars.
Though \textit{Gaia} does detect white dwarfs \citep[e.g.][]{GentileFusillo_2021,Jimenez-Esteban_2023}, it does not measure their age or composition, so for consistency we do not include them here. 

\textit{Gaia}, like any survey, has selection effects: it does not observe all stars, does not make every measurement for each of the stars it observes, and does not choose stars to make measurements for in an unbiased way.
Thus in order to reconstruct the true underlying stellar population from our subsample, our subsample needs to be debiased.

Firstly, we estimate the subsample selection function (SSF): the fraction of stars in the underlying stellar population that make it into the subsample we define above, as a function of color, magnitude and on-sky position.
This is equal to the fraction of the stars at each colour, magnitude and on-sky position in the \textit{Gaia} catalogue that is also in our subsample, multiplied by the catalogue completeness. The catalogue completeness is the fraction of the underlying population that makes it into the \textit{Gaia} catalogue (i.e. has at least one measurement recorded) however \cite{Cantat-Gaudin_2023} find that for magnitudes brighter than the limit of our subsample (\(G\lesssim14\)) the catalogue is \(100\%\) complete, so we do not need to calculate this. This means we can estimate SSF following the method of \cite{Castro-Ginard_2023}.
\cite{Castro-Ginard_2023} find that the selection function for the subsample of stars with a radial velocity measurement depends on \(G\) magnitude, \(G-G_\text{RP}\) color and on-sky position.
Since the additional measurements we require (GSP-Spec \(\MH\) and GSP-Spec-based FLAME age) are both computed from the output of the RVS, we follow \cite{Castro-Ginard_2023} in assuming that our SSF depends on magnitude, color and on-sky position.
The method of \cite{Castro-Ginard_2023} calls for counting the number of stars in the subsample \(k\) and number of stars in catalogue \(n\) in bins in magnitude, color and on-sky position \citep[using HEALPix\footnote{Hierarchical Equal Area isoLatitude Pixelation};][]{Gorski_2005}. 
The SSF is then estimated in each bin as \((k+1)/(n+2)\) with an uncertainty of \(\sqrt{\frac{(k+1)(n-k+1)}{(n+2)^2(n+3)}}\).
We define our subsample as having measurements of radial velocity, \(\MH\) and age with the same cuts on errors and flags as our data described above; the exact query to the \textit{Gaia} archive used is listed in Appendix~\ref{sec:gaiaArchiveQueries}.
We calculate the SSF in bins in color and magnitude of width \(\Delta G=1\) and \(\Delta(G-G_\text{RP})=0.2\).
Following \cite{Castro-Ginard_2023} we group the Galactic polar caps (Galactic latitudes \(\abs{b}>\qty{30}{\deg}\)) and bin the rest of the sky in HEALPix level 2, a level that retains the smooth shifts in the selection function. 
These bins have a larger width in magnitude and on-sky position than the binning of \cite{Castro-Ginard_2023}; as our subsample is smaller, requiring more measured quantities, we use broader bins to retain statistical quality. 
To ensure uniformity across the sky and a low fractional uncertainty in the effective selection function described below, we set the SSF equal to zero in all colour-magnitude bins with any on-sky bin where \(k<2\).
This does not affect any color-magnitude bins where a significant number of stars are observed, retaining 98\% of the stars in the subsample.
This gives us the selection function in magnitude, colour and sky position.

Secondly, we estimate the effective selection function (ESF): the fraction of stars in the underlying stellar population that make it into the subsample we define above, as a function of their metallicity, age, distance, and on-sky position. We estimate this by combining the SSF, a function of colour and magnitude, with PARSEC isochrones \citep{Bressan_2012, Marigo_2017}, stellar models which predicts the colour and magnitude stars with a given metallicity, age and distance have at each mass.
We estimate the ESF by integrating over mass the SSF weighted by a Kroupa IMF \citep{Kroupa_2001} along PARSEC isochrones, calculating the fraction of the modelled stellar population that would make it into our subsample at each metallicity, age, distance, and on-sky position.

The reason we need to use the ESF to debias our sample is that the SSF is non-zero only in a narrow colour range (\(0.2\leq G-G_{RP}<0.6\)), so we cannot account for the contributions of stars outside of this range directly.
Conversely, the ESF remains non-zero across almost all of parameter space, meaning the entire stellar population within \qty{200}{\parsec} can be reconstructed.

We calculate the ESF using PARSEC isochrones at \(\MH\) between -1.45 and 0.65 spaced by 0.1 and ages from 0.5~Gyr to 13.5~Gyr spaced by 1~Gyr, and the universal initial mass function of \cite{Kroupa_2001}.
When converting between absolute and apparent magnitudes we do not account for dust extinction, as within 200~pc from the Sun this is limited to 1~mag at most in the G band \citep{Green_2019, Lallement_2022}.

Stars can be assumed to form with a consistent initial mass function \citep{Chabrier_2001,Kroupa_2013}, but to quantify how varying the IMF used affects our results, we also calculate an alternative ESF using the log-normal IMF of \cite{Chabrier_2003}.
Since our results consist of relative distributions and not absolute values, to compare the two ESFs we scale them to have the same mean value, then calculate the mean fractional difference between the two to be only \(0.8\%\).
Thus the choice of IMF does not affect our results, and we use the Kroupa-based ESF for the rest of this work.

The final effect we account for is stellar death.
Stars born throughout Galactic history emit ISOs, and we expect ISOs to outlive their parent stars: for instance, \(\gtrsim\)\qty{10}{\kilo\meter}-diameter Oort cloud comets have survived \qty{4.5}{\giga\year} in an interstellar environment with little erosion \citep{Guilbert-Lepoutre_2015}, so similar-scale ISOs will largely remain intact.
This means we cannot use the distribution \textit{Gaia} observes of only the currently-living stars, or the proportion of young ISOs would be overestimated: Fig.~\ref{fig:stars} (Age) shows the difference between the age distributions of the raw Gaia data points and the reconstructed sine morte distribution.
Instead, we must estimate what the stellar population would be without stellar death: the \textit{sine morte} stellar population.
We account for this in our model at the same time that we account for \textit{Gaia}'s selection effects:
we define the ESF as the fraction of all stars ever formed (i.e. the fraction of the \textit{sine morte} population) at a given age, metallicity, on-sky position and distance that would be observed. 

Unlike the SSF, the ESF is nonzero for almost all of its parameter space, meaning all populations are represented by at least a fraction of their stars in the \textit{Gaia} sample and thus their contributions can be accounted for.
The only exceptions to this are in bins at distances closer than \qty{25}{\parsec}, so to ensure uniformity we cut the few stars with observed parallax \(>\qty{40}{\milli\arcsecond}\), leaving a final data sample of size \num{201426}. 

We list the ADQL queries made to the Gaia archive to retrieve all datasets in Appendix~\ref{sec:gaiaArchiveQueries}.

\subsection{The Local \textit{Sine Morte} Stellar Distribution}
\label{sec:sine_morte}

\begin{figure}[t]
\centering
\includegraphics[width=\figwidth\textwidth]{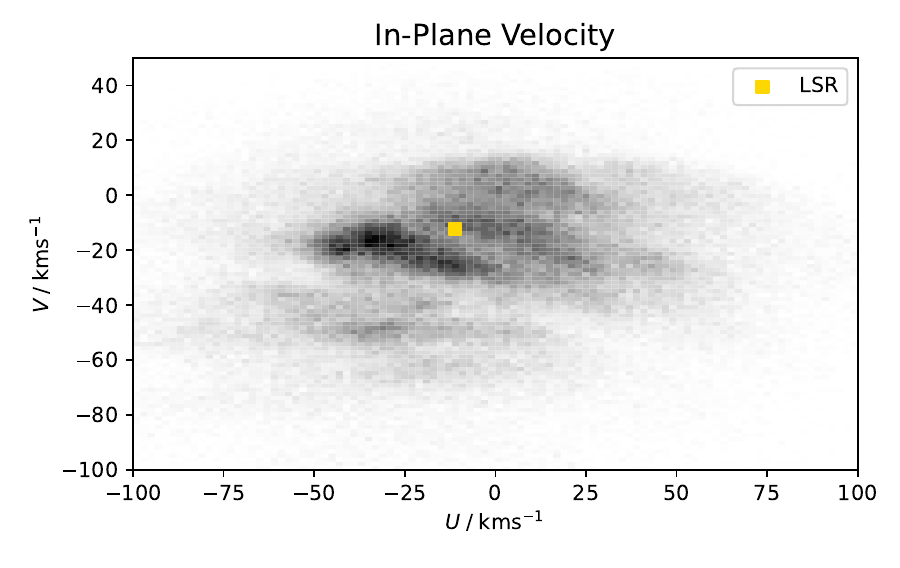}
\includegraphics[width=\figwidth\textwidth]{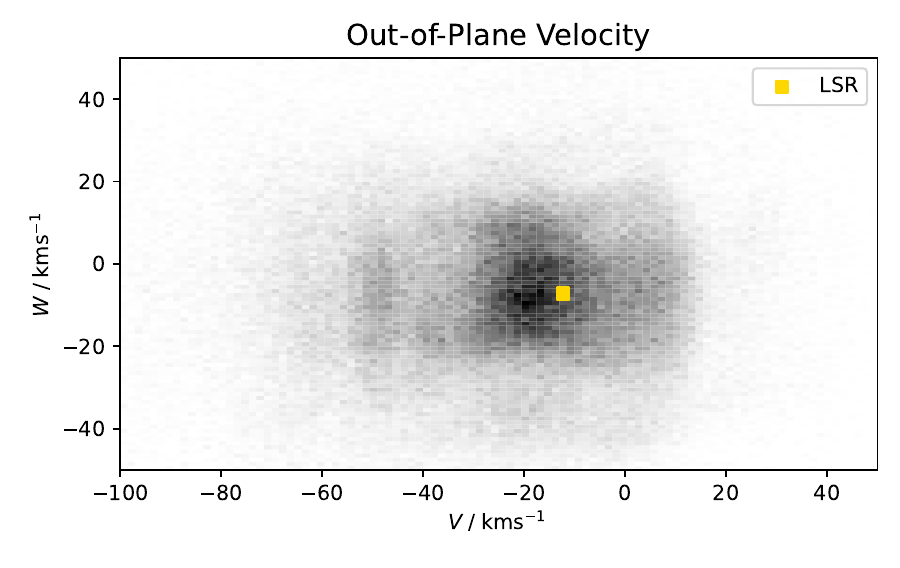}

\includegraphics[width=\figwidth\textwidth]{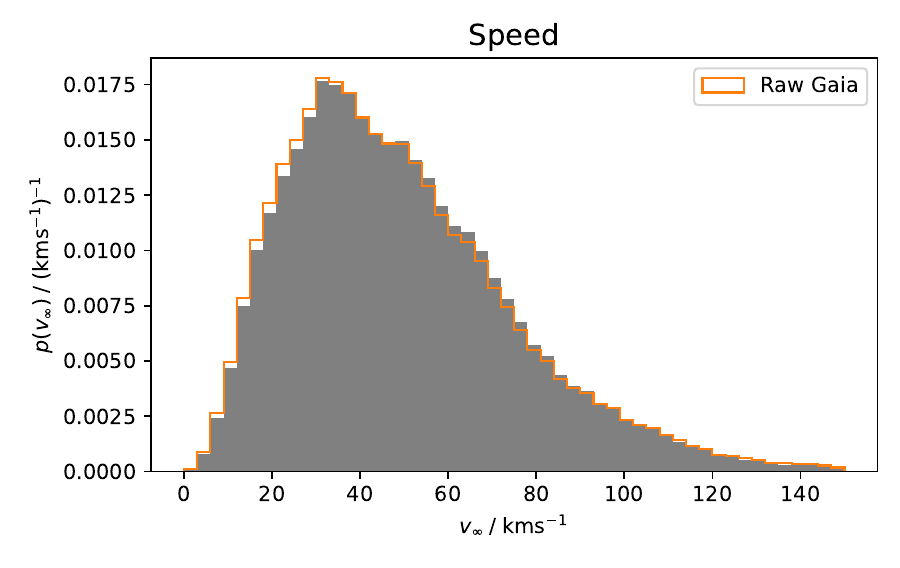}
\includegraphics[width=\figwidth\textwidth]{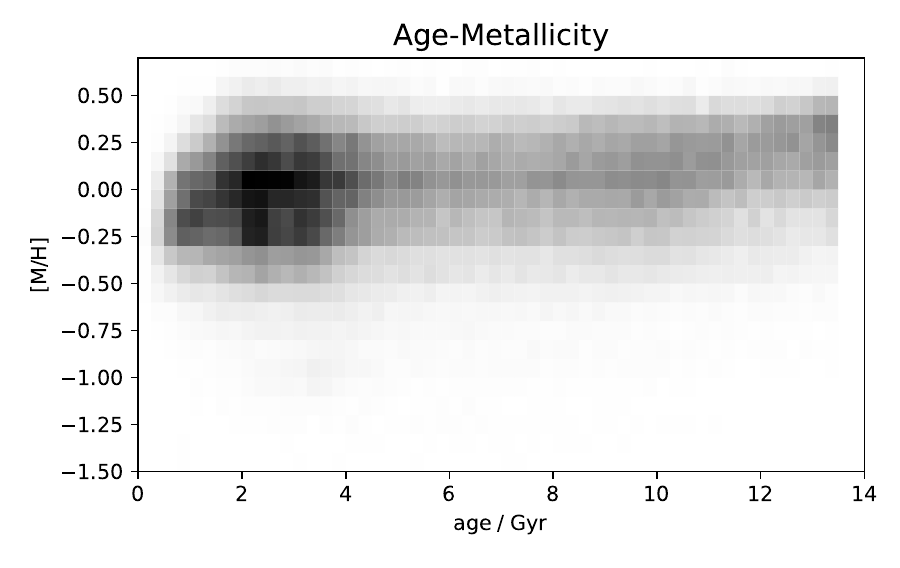}

\includegraphics[width=\figwidth\textwidth]{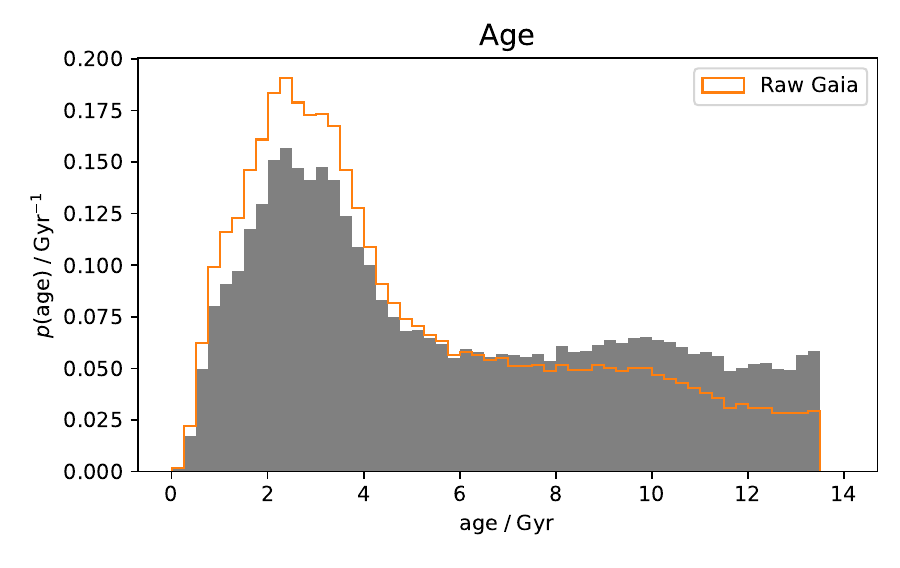}
\includegraphics[width=\figwidth\textwidth]{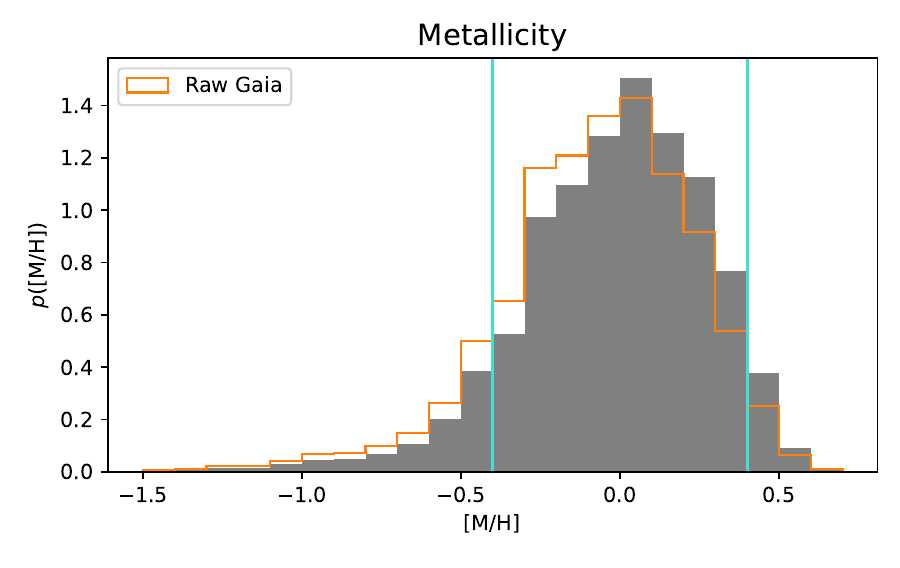}
\caption{Distributions of the \textit{sine morte} stellar population in velocity, age and \(\MH\). \(U\) is the component of velocity towards the Galactic centre, \(V\) in the direction of motion around the Galactic centre, and \(W\) out of the plane of the Galaxy, each relative to the Sun's velocity. Both 1D and 2D histograms are normalised to integrate to unity. In orange unfilled histograms we have also plotted the distribution of the raw \textit{Gaia} sample, without accounting for selection effects or stellar death, for comparison. The vertical lines on the \(\MH\) distribution are the limits of the PPD chemical model.}
\label{fig:stars}
\end{figure}

In the previous section we described how we acquired a sample of stars from \textit{Gaia} with a set of observed properties, and defined the ESF: the fraction of the underlying \textit{sine morte} stellar population included in our sample as a function of their observed properties.
This allows us to debias our observed sample and estimate the underlying distribution of these properties in the \textit{sine morte} stellar population.
We do this by treating our sample of stars as a weighted sampling of the \textit{sine morte} population, weighted by the inverse of the value of the ESF for each star.
Since the ESF is the fraction of \textit{sine morte} stars included in our sample, its inverse gives the number of stars in the underlying \textit{sine morte} population represented by each observed star.
The median value of the ESF over the stars in our sample is 1/22.5, meaning this median observed star represents 22.5 stars in the \textit{sine morte} population.
This gives us the Solar neighbourhood \textit{sine morte} stellar distribution in velocity, age and metallicity \(\MH\), plotted in Figure~\ref{fig:stars}.

In Fig.~\ref{fig:stars} (In-Plane Velocity and Out-of-Plane Velocity) the velocity distribution is plotted in its right-handed Cartesian components relative to the Sun's velocity: \(U\) towards the Galactic Centre, \(V\) in the direction of Galactic rotation, and \(W\) out of the plane of the Galaxy.
Immediately obvious is the rich structure in the stellar velocity distribution: the moving groups and branches \citep[compare to Fig.~5 of][for identification]{Antoja_2008}. 
As described in the introduction, these overdensities are caused by resonances with the Galactic spiral arms and bar carving out furrows in the velocity distribution, and are not dispersed clusters.
There is less structure in the \(W\) distribution than the \(U\) and \(V\) distribution.
With a yellow square we mark the local standard of rest \citep[LSR;][]{Schonrich_2010}, the velocity that a star on a circular orbit would have in a azimuthally-smoothed Galactic potential, which approximates the actual potential. 

We measure an age distribution for the Solar neighborhood plotted in Fig.~\ref{fig:stars} (Age) that is similar to that of \cite{Nordstrom_2004}. 
We also recover a known feature of the stellar population of the Solar neighbourhood: a flat age-metallicity relation, with a large scatter in metallicity at all ages, meaning stars of all metallicities are found at all ages (Fig.~\ref{fig:stars}, Age-Metallicity).
The feature is surprising, as one would expect stellar nucleosynthesis to cause the metallicity of stars to increase over time and thus decrease with age, however it is attested in many surveys of stars \citep{Edvardsson_1993,Haywood_2013} as well as surveys of white dwarfs \citep{Rebassa-Mansergas_2021}.
The reason for the Solar neighbourhood's flat age-metallicity relation is debated, with explanations including an inside-out formation of the Galactic disk with radial migration \citep{Sellwood_2002, Schonrich_2009}, and 
a two-phase formation of the Galactic disk \citep{Haywood_2013}.

\subsection{Stars Produce Interstellar Objects}
\label{sec:stars_to_isos}

To predict the Solar neighbourhood ISO distribution from the \textit{sine morte} stellar distribution, we follow a similar method to \citet{Hopkins_2023}. 
As in \citet{Hopkins_2023} Sec. 3.1, we map stellar metallicities to ISO compositions.
The model of \citet{Bitsch_2020} generates a clear trend for a planetesimal's water mass fraction \(\fHHO\) with stellar metallicity, for planetesimals beyond the water ice line (see their Fig. 10).
We expect the majority of ISOs to originate in this region due to the greater number of planetesimals and the higher efficiencies of ejection mechanisms at larger distances from the central star \citep{Fitzsimmons_2023}.
Though the relation in \cite{Bitsch_2020} correlates \(\fHHO\) with \(\FeH\), the relative abundance of iron specifically, this is well approximated by \textit{Gaia}'s overall metallicity \(\MH\) for the thin-disk, non-alpha-enhanced majority of stars in the Solar neighbourhood \citep{Salaris_2005}.
We therefore use this water mass fraction as our defining compositional property for the chemodynamic analysis.
\cite{Bitsch_2020} predict that the water mass fraction of planetesimals decreases with increasing metallicity. 
This chemical model is limited to metallicities within \(-0.4\leq\MH\leq0.4\), corresponding to \(0.07\leq\fHHO\leq0.51\); the majority of our ISOs do lie within these limits.
Outside of this range, we assume that the relationship between \(\fHHO\) and \(\MH\) continues to be monotonic, with \(\fHHO\) remaining high beyond the low \(\MH\) limit and remaining low beyond the high \(\MH\) limit.

Though we assume here that the majority of ISOs form beyond the water ice line, there have been several mechanisms suggested which may eject planetesimals from much closer to their parent star \citep[e.g.][]{Cuk_2018,Rafikov_2018,Childs_2022}. 
These ISOs would be chemically distinct as they would most likely be devolatilised.
Future modelling should consider the likely contribution of these populations to the observed ISO fraction, but will need detailed consideration of non-linear processes within the disk as well as a more sophisticated chemical treatment.
However, we note that both discovered ISOs so far appear to originate from beyond the H$_2$O ice line, with the non-gravitational acceleration of 1I suggesting the presence of volatiles \citep{Micheli_2018} and 2I originating from beyond the CO ice line \citep{Seligman_2022}.
Additionally, even for planetesimals formed beyond the water ice line, the one-to-one relation between \(\fHHO\) and \(\MH\) in \cite{Bitsch_2020} is a simplification, as in reality stars of one metallicity will produce ISOs with a range of compositions.
\cite{Cabral_2023} show that though the exact mass fraction values in the relation are uncertain, they confirm the trend of decreasing \(\fHHO\) with \(\FeH\) to be robust.

Since the mass of planetesimal-forming metals in a protoplanetary disk appears proportional to the metal fraction of its central star \citep[e.g.][]{Lu_2020}, we assume that the number of ISOs produced by a star of a given metallicity \(\MH\) is proportional to \(10^{\MH}\).
We explored the sensitivity of the ISO population to the metallicity-dependence relationship in \citet{Hopkins_2023};
that work's Sec.~4.2 showed that if there were no metallicity dependence, more high-\(\fHHO\) ISOs would be present.

To link the velocity distribution of ISOs to that of stars, the following points of dynamics need to be considered.
Firstly, we should expect the ISO velocity distribution to have the same pattern of gaps and overdensities as the stellar velocity distribution, without any assumptions about their origins.
As discussed in the introduction, the structures in the stellar velocity distribution are caused by resonances with the spiral arms and bar, compared to which individual ISOs and stars both have negligible mass.
This means the Galactic orbits of ISOs will be affected by these resonances to the same extent as the orbits of stars.
These structures are being actively formed, and since the potentials which cause these resonances are transient with a timescale of \(\sim\qty{100}{\mega\year}\) \citep{Baba_2015} the velocity distribution structure must form on this timescale or less.
This is significantly less than the average age of stars and ISOs in the Solar neighbourhood, thus the ISO velocity distribution must develop the same pattern of gaps and overdensities as the stellar velocity distribution, regardless of their initial velocity distribution.

Secondly, ISOs do not stay near their parent stars.
The vast majority of stars in the Solar neighbourhood belong to the Milky Way disk so are on near-circular orbits around the Galactic centre with an orbital speed of \(\sim\qty{200}{\kilo\meter\per\second}\) \citep{Bovy_2012}.
All significant mechanisms to eject planetesimals from their parent planetary systems produce ISOs with ejection velocities significantly less than this orbital velocity \citep[\(<\qty{10}{\kilo\meter\per\second}\)][]{Hands_2019,Pfalzner_2021}, therefore ejection puts ISOs onto similar orbits to their parent stars.
However, these small differences in velocity mean that ISOs do not stay near their parent star, quickly dispersing along this shared near-circular orbit, azimuthally. 
This means that most of the ISOs we discover passing through the inner Solar system do not have parent stars in the immediate Solar neighbourhood, nor in our \textit{Gaia} stellar sample.

Regardless, we can still predict the Solar neighbourhood ISO distribution from the Solar neighbourhood stellar distribution, because due to their velocity dispersion stars mix azimuthally too\footnote{This is very well demonstrated by the video showing the orbits of a selection of stars from the Gaia Catalogue of Nearby Stars at \url{https://www.cosmos.esa.int/web/gaia/edr3-gcns}, or equivalently Figure~21 of \cite{GaiaCollaboration_2021}.}.
Under the assumption that the Galaxy does not have large deviations from axisymmetry, the distribution of the whole ISO population will follow the distribution of whole population of stars.
Since stars and ISOs are equally negligible in mass compared to the spiral arms, bar and giant molecular clouds \citep{Gustafsson_2016} that dominate the evolution of their orbits, the orbits of ISOs still evolve in the same way as those of stars.\footnote{N.B. close encounters with individual stars don't contribute to the evolution of the orbits of either stars or ISOs as these are incredibly rare: stars and ISOs will undergo the same rate of close encounters with individual stars, and the timescale for an object with a speed of \qty{55}{\kilo\meter\per\second} undergoing a \qty{5}{\astronomicalunit}-separation encounter with stars of density \qty{0.1}{\per\parsec\cubed} is \qty{1e5}{\giga\year}. For a population of age \(\sim\qty{10}{\giga\year}\), we would expect only 0.01\% of ISOs (and stars) to have undergone such a close encounter.}

In summary, though individual ISOs do not stay near their own parent stars, the distribution of the whole ISO population follows the distribution of the whole stellar population. Applying the metallicity dependence detailed above gives
\begin{equation}\label{eq:stars2ISOs}
n_\text{ISO}(\mathbf{x},\mathbf{v},\tau,\fHHO)\,\mathrm{d}\fHHO \propto 10^{\MH} \cdot n_\text{stars}(\mathbf{x},\mathbf{v},\tau,\MH)\,\mathrm{d}\MH
\end{equation}
where \(n_\text{ISO}\) and \(n_\text{stars}\) are the number density distributions of ISOs and stars respectively in position \(\mathbf{x}\), velocity \(\mathbf{v}=(U,V,W)\), age \(\tau\) and composition as measured by \(\MH\) for stars and \(\fHHO\) for ISOs. \(\fHHO\) and \(\MH\) are linked by the one-to-one relation discussed above, and we treat these distributions as being constant in \(\mathbf{x}\) over our Solar neighbourhood sample and focus on the other variables.

We expect the vast majority of ISOs to be released within several hundred Myr of a star’s
formation \citep{Pfalzner_2019, Lisse_2022, Fitzsimmons_2023}. Since this is significantly less than the average age of stars in the Solar neighbourhood, we do not account for a substantial delay between a star's birth and the release of its ISOs.
Finally, we expect ISOs to outlive their parent stars (see \S~\ref{sec:SF}).

The stars in the observed \textit{Gaia} subsample with measured velocity, metallicity and age can be thought of as a biased sampling from the chemodynamical distribution shared by both stars and ISOs.
Just as weighting these samples by the inverse of the effective selection function debiases the sample and gives us a prediction of the \textit{sine morte} stellar distribution, to predict the distribution of ISOs in the Solar neighbourhood we reweight these samples by the additional ISO production factor of \(10^{\MH}\), following Eq.~\ref{eq:stars2ISOs}.
This distribution is plotted in Fig.~\ref{fig:underlyingISO} in Appendix~\ref{sec:grav_focussing_appendix}.
The distribution of ISOs in the Solar neighbourhood however is not the distribution of ISOs that is observable from our place in the inner Solar system.

\subsection{Within the Solar System: The Volume Sampling Rate and Gravitational Focussing}
\label{sec:grav_focussing}
ISO-detecting surveys such as Pan-STARRS \citep{Chambers_2016} and LSST have run-times significantly longer than the typical lengths of time ISOs are visible: 1I and 2I spent only 2 months and 3 years respectively brighter than 24 mag, the LSST single-exposure \(r\)-band point source depth \citep{Ivezic_2019}.
Thus the population of ISOs observable at any time is constantly being refreshed.
This refreshing effect \citep[previously discussed in][]{Stern_1990,Moro-Martin_2009} means that the distribution of ISOs that are observable is the distribution of those streaming through the inner Solar system, not the distribution of the static population in the Solar neighbourhood of \S~\ref{sec:stars_to_isos}.
This streaming population experiences two velocity-dependent effects that change their distribution in parameter space relative to the static population.
First is the dependence of this refresh rate on relative speed \(v_\infty=\sqrt{U^2+V^2+W^2}\): for the same spatial density, the observable volume samples ISOs with high \(v_\infty\) at a proportionally higher rate.
The second effect is gravitational focussing: the increase of the cross section \(\sigma\) at low \(v_\infty\) for encounters with ISOs with perihelion less than \(q\) \citep{Whipple_1975,Seligman_2018,Forbes_2019}.\footnote{This treatment of gravitational focussing differs from that of the \textit{probabilistic method} of \cite{Marceta_2023b} who calculates the effect on the distribution of ISOs within the Solar system at a single moment in time, as opposed to the streaming population.}

\begin{figure}[t]
    \centering
    \includegraphics[width=\figwidth\textwidth]{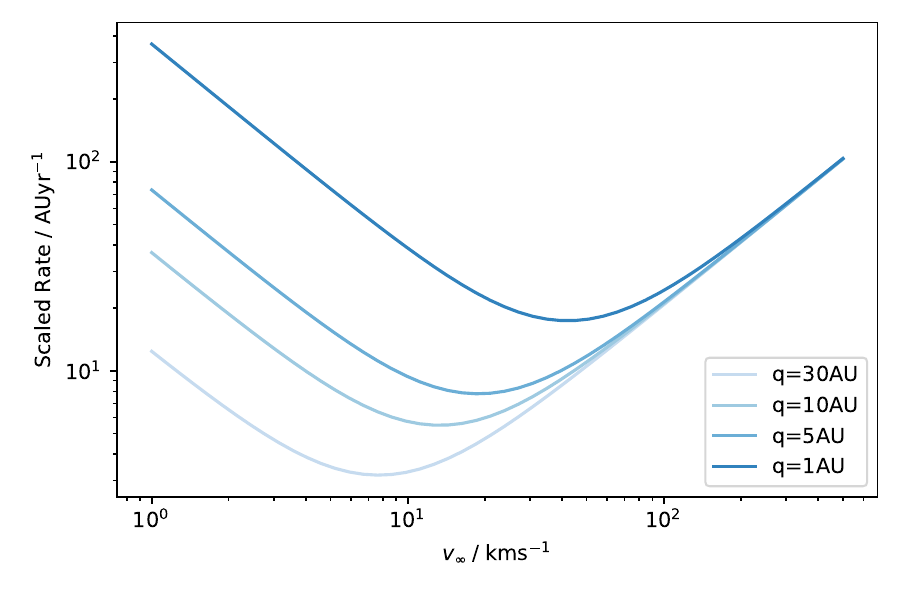}
    \caption{Volume sampling rate scaled by \(1/\pi q^2\) for different perihelia \(q\). This is the combination of gravitational focussing, which increases the sampling rate of low relative speed ISOs, and the ``refresh rate'' effect, where faster ISOs are sampled at proportionally higher rates.
    }
    \label{fig:volsamprate}
\end{figure}

These effects are combined in the volume sampling rate:
\begin{equation}\label{eq:VSR}
\gamma(q,v_\infty) = v_\infty \sigma = v_\infty \pi q^2 \Bigg(1+\frac{2GM_\odot}{q v_\infty^2}\Bigg)  
\end{equation}
This follows from classical dynamics and is first referenced in Eq.~5 of \cite{Whipple_1975}.
The volume sampling rate is plotted in Fig.~\ref{fig:volsamprate}, scaled by \(1/\pi q^2\), the inverse of the geometric cross section.
Multiplying the velocity distribution of ISOs in the Solar neighbourhood by \(\gamma(q,v_\infty)\) gives the rate of ISOs with perihelion less than \(q\) entering the inner Solar system.

Though the volume sampling rate becomes proportional to \(v_\infty^{-1}\) at low speeds, it is unnecessary to apply a cutoff or gravitational softening.
This is because the the velocity distribution remains integrable due to the \(v_\infty^2\) Jacobian factor that arises from the velocity distribution being three-dimensional.
Thus the one-dimensional speed distribution (Fig.~\ref{fig:isoq5}, Speed) becomes proportional to \(v_\infty\) at small speeds, and does not diverge.
We find adding a gravitational softening parameter \(\phi=\qty{2}{\kilo\meter\per\second}\) \citep[as defined in][]{Seligman_2018} does not change the likelihoods calculated in \S~\ref{sec:modelcomparison}.

To predict the distribution of ISOs entering the inner Solar system, we reweight the Solar neighbourhood ISO distribution described in \S~\ref{sec:stars_to_isos} by the volume sampling rate for a given \(q\), here \(q=\qty{5}{\astronomicalunit}\).
The weights applied to the raw \textit{Gaia} samples to get each distribution is summarised in Table~\ref{tab:weights}.

\begin{table}[h]
\hspace{-1.1cm}\begin{tabular}{c|c c c c}
\hline
Distribution \textcolor{white}{.}& Raw \textit{Gaia} Samples \textcolor{white}{.}& \textit{Sine Morte} Stars \textcolor{white}{.}& Solar Neighbourhood ISOs \textcolor{white}{.}& \(q<\qty{5}{\astronomicalunit}\) ISOs \\[3pt]
\hline
&&&&\\[-7pt]
Weight & 1 & \(\dfrac{1}{\text{ESF}}\) & \(\dfrac{10^{\MH}}{\text{ESF}}\) &  \(\dfrac{10^{\MH}\cdot\gamma(\qty{5}{\astronomicalunit},v_\infty)}{\text{ESF}}\) \\[10pt]
\hline
\end{tabular}
\caption{Summary of weights given to observed Gaia data points to reconstruct the stellar and ISO distributions used. The effective selection function (ESF) is a function of each \textit{Gaia} sample's distance, age, metallicity and on-sky position.}
\label{tab:weights}
\end{table}

\subsection{Known Objects}
\label{sec:known}

Our comparison sample is the two currently known\footnote{While the prospect of interstellar meteor detection has been long-standing for many decades, we rely on the assessment of \citet{Brown_2023} that no interstellar meteoroids are yet confirmed.} macroscopic interstellar objects: 1I/\Ou\ and 2I/Borisov.
Objects within the Solar System are defined as interstellar based on their orbital energy.
Fitting to measured astrometry gives the derived value of the original reciprocal of the semimajor axis at a given distance from the Sun, 1/$a_{ori}$, which must be dominantly negative to be hyperbolic (giving an eccentricity $e\gg1$).
This gives the asymptotic speed or velocity at infinity, \(v_\infty=\sqrt{U^2+V^2+W^2}\): the speed of an ISO on its hyperbolic orbit relative to the Sun before entering the Solar System. 
Both known ISOs were measured for sufficiently long arcs, and relative to \textit{Gaia}-level stellar catalogues, that the precision of their original orbits is well constrained.
We use the barycentric original orbit at a distance of \qty{250}{\astronomicalunit} from the Sun of \citet{Bailer-Jones_2018} and \citet{Bailer-Jones_2020} for 1I and 2I. 
It is possible that \citet{Krolikowska_2023}'s recent finding of non-gravitational acceleration in pre-perihelion comets at \(r_h > \qty{5}{\astronomicalunit}\) may mean a future revision of 2I's original orbit, and introduces some degree of uncertainty to 1I/\Ou's original orbit; this seems relatively unlikely at present, given the variations in NG accelerations tested by various authors\footnote{e.g. Table 1 in both \citet{Bailer-Jones_2018,Bailer-Jones_2020}, respectively; also \cite{Dybczynski_2018} and \cite{Dybczynski_2019}, \citet{Micheli_2018}.}.
Unlike 2I, 1I/\Ou's orbit was not precovered into pre-perihelion observations, so only its post-perihelion non-gravitational (NG) accelerations are quantified \citep{Micheli_2018}.
Nevertheless, the known velocities of both ISOs are uncertain only to tens of metres per second, with the speed and direction changing by factors of only 1 in 1000 both between NG acceleration models and within the uncertainties of each model --- and it is the velocities that are the key parameter for our analysis here. 
For the post-encounter asymptotic velocity, we use the full observed arc orbits of JPL Horizons\footnote{Solar System Dynamics. (Downloaded 2023-12-20). 1I: \url{https://ssd.jpl.nasa.gov/tools/sbdb_lookup.html\#/?sstr=1I}, solution as of 2018-Jun-26 12:17:57; 2I: \url{https://ssd.jpl.nasa.gov/tools/sbdb_lookup.html\#/?sstr=2I}, solution as of 2020-Aug-21 09:32:58.}.

Some Solar System comets have been considered as potential marginal cases of interstellar objects, often with weakly hyperbolic $e\sim1.01-1.05$.
For reference of our distributions against such long-period comets, we use as an example the bright comet C/1956 R1 (Arend–Roland), which was at one time considered as potentially the first interstellar comet \citep{Sekanina_1968}.
It has a 497-day arc (with astrometry measured relative to non-\textit{Gaia}-calibrated photographic plates), and is among a small group of comets with a high N$_2$/CO$_2$ production ratio \citep{Anderson_2023}.

\section{Results}
\label{sec:results}

\subsection{Interstellar Object Distribution}\label{sec:theDist}

\begin{figure}[p]
\centering
\includegraphics[width=\figwidth\textwidth]{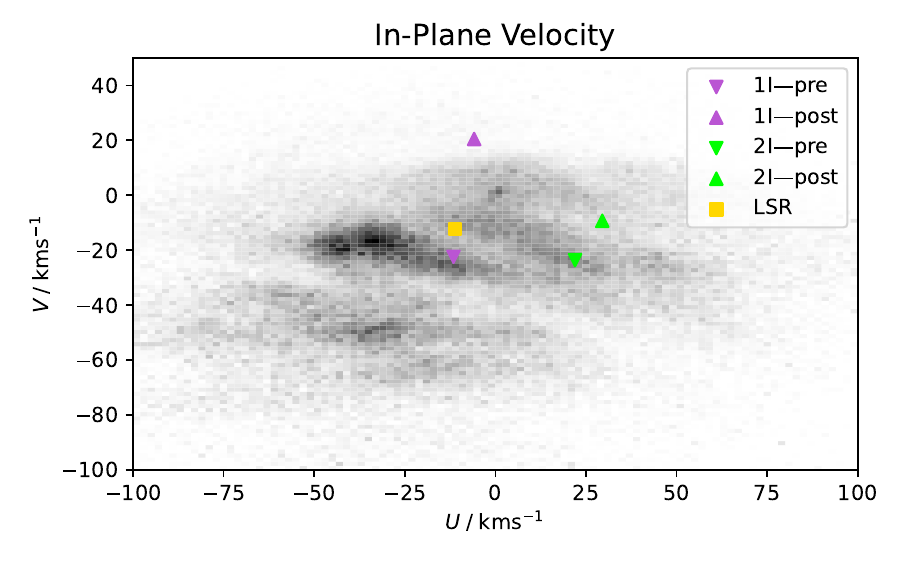}
\includegraphics[width=\figwidth\textwidth]{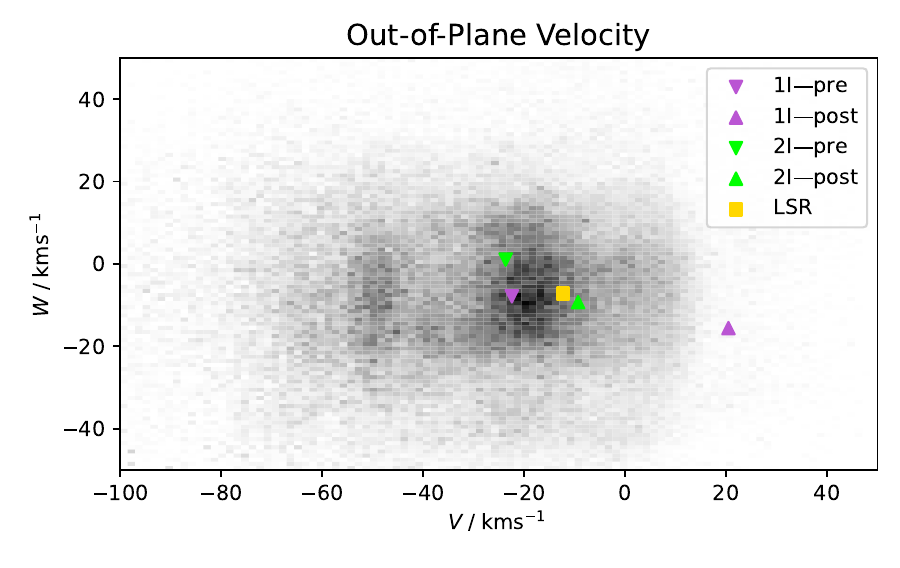}

\includegraphics[width=\figwidth\textwidth]{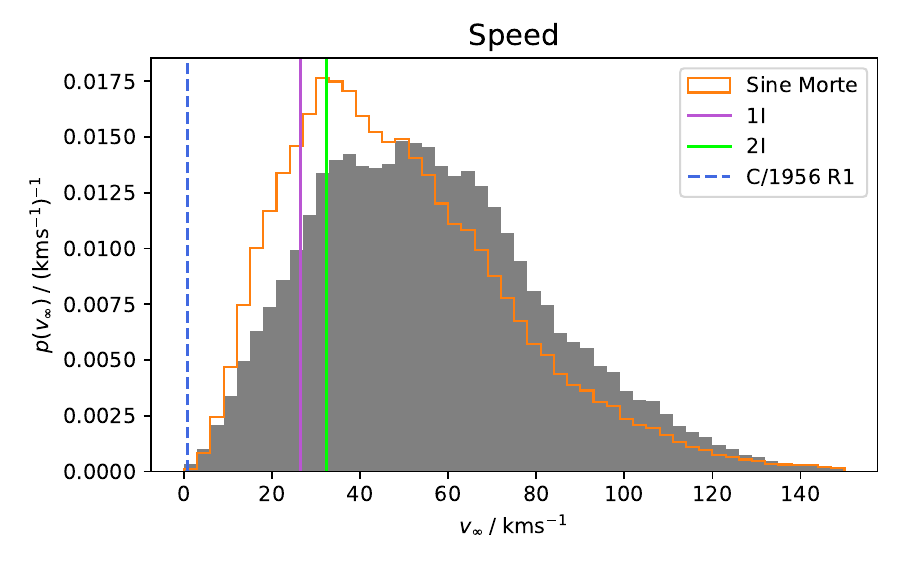}
\includegraphics[width=\figwidth\textwidth]{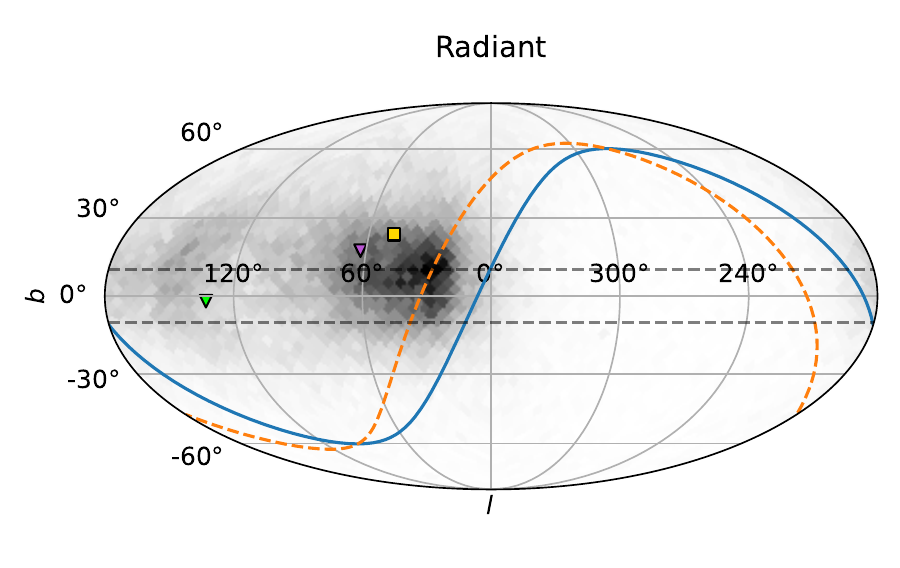}

\includegraphics[width=\figwidth\textwidth]{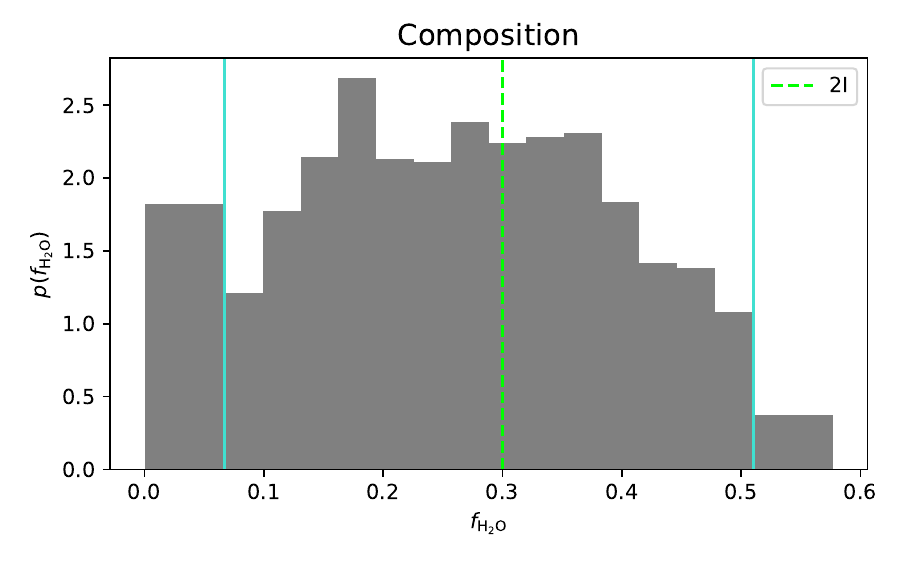}
\includegraphics[width=\figwidth\textwidth]{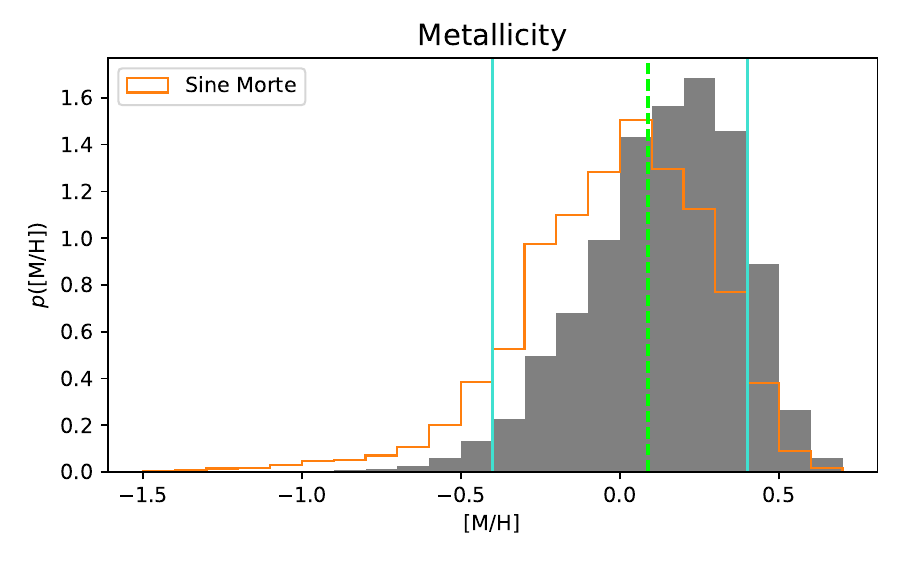}

\includegraphics[width=\figwidth\textwidth]{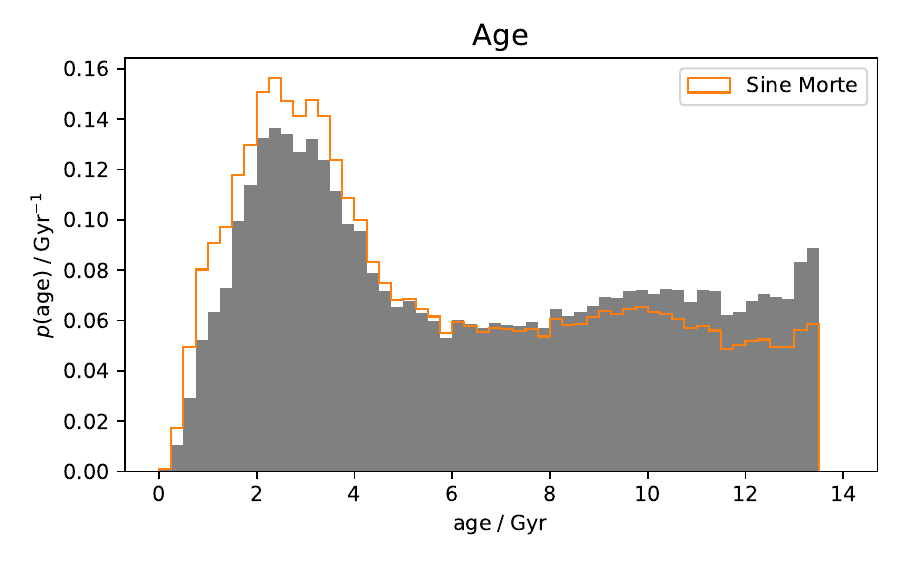}

\caption{Predicted distributions of ISOs with \(q<\qty{5}{\astronomicalunit}\) in velocity, on-sky radiant, composition proxied by water mass fraction \(\fHHO\), parent-star metallicity and age, analogous to Fig.~\ref{fig:stars}. On the radiant plot the ecliptic plane is marked in solid blue, the celestial equator is marked in dashed orange, and dashed horizontal lines encapsulate the $b=\pm10^{\circ}$ region of high stellar density along the Galactic plane. Orange unfilled histograms plot the distribution of the \textit{sine morte} stellar population for comparison.}
\label{fig:isoq5}
\end{figure}

Plotted in Figure~\ref{fig:isoq5} is the predicted distribution of \(q\leq\qty{5}{\astronomicalunit}\) ISOs entering the Solar System in two-dimensional velocity, speed, on-sky radiant, composition, source stellar metallicity and age.

We first consider the velocity distribution of ISOs in their Cartesian components \(U\), \(V\) and \(W\) in two two-dimensional histograms (Figure~\ref{fig:isoq5}, In-Plane Velocity and Out-of-Plane Velocity).
As expected given that stars are the progenitors, under our model in \S~\ref{sec:stars_to_isos} the ISO velocity distribution shows the same moving groups and branches as the stellar distribution of Figure~\ref{fig:stars}.
Many ISOs will have pre-encounter velocities in these structures: we predict 18\% of ISOs passing through the inner Solar system will originate in the largest of these, the Hyades-Pleiades branch\footnote{defined here with boundaries \(V<\qty{-5}{\kilo\meter\per\second}\), \(U>\qty{-55}{\kilo\meter\per\second}\), \(V>\qty{-35}{\kilo\meter\per\second}-0.2U\), and \(V<\qty{-25}{\kilo\meter\per\second}-0.6U\), following Fig.~5 of \cite{Antoja_2008}.}. 
However, the ISOs display an additional peak at the Sun's velocity (\(U,V,W = 0,0,0\)) when compared to the stellar distribution due to gravitational focussing, discussed in \S~\ref{sec:grav_focussing} and illustrated in Fig.~\ref{fig:volsamprate}. 
We use a $q \leq 5$~au cut as indicative for \frenchspacing{e.g.} ease of observational characterisation, as typical of Solar System comets. 
For comparison, the resulting ISO distributions with $q$ cuts instead made at \qty{1}{\astronomicalunit} and at \qty{30}{\astronomicalunit} are shown in Appendix~\ref{sec:grav_focussing_appendix}.
The Solar velocity concentration becomes more pronounced as the perihelion cut decreases, as expected; by $q = 30$~au, there is effectively no gravitational focussing.
As in Fig.~\ref{fig:stars} we mark with a yellow square the local standard of rest \citep[LSR;][]{Schonrich_2010}, the indicative velocity of a star on a circular orbit in a theoretical azimuthally-smoothed Galactic potential. 

To compare this predicted population with the known sample of detected interstellar objects, we indicate in Figure~\ref{fig:isoq5} the Cartesian component velocities of 1I/\Ou\ and 2I/Borisov both before and after their perturbing interaction with the Solar System (\S~\ref{sec:known}): 1I is marked in purple and 2I in green.
The pre-encounter velocities are directly comparable to our predicted population; the post-encounter velocities are to illustrate the effects of the only two measured ISO gravitational interactions with the Sun.
Both known ISOs' pre-encounter velocities are entirely typical for the population, occurring in regions of velocity space with relative ISO abundance.
This places both ISOs within moving groups in velocity space; we discuss the implications of this in \S~\ref{sec:origins}.

We next consider the asymptotic speed $v_\infty$  distribution of ISOs (Figure~\ref{fig:isoq5}, Speed).
This asymptotic speed distribution is similar to that predicted by \cite{Eubanks_2021}, but weighted higher, with our median asymptotic speed being \qty{55}{\kilo\meter\per\second}, compared to their \qty{38}{\kilo\meter\per\second}.
This is largely due to our choice in these plots of calculating gravitational focussing for \(q<\qty{5}{\astronomicalunit}\) rather than \cite{Eubanks_2021}'s \(q<\qty{1}{\astronomicalunit}\), which places less weighting on low-speed ISOs.
When we calculate the speed distribution for \(q<\qty{1}{\astronomicalunit}\) (Fig.~\ref{fig:isoq1}, Speed) we get a median speed of \qty{45}{\kilo\meter\per\second}.
For large perihelia the effect of gravitational focussing becomes negligible (e.g. for \(q<\qty{30}{\astronomicalunit}\) ISOs, plotted in Fig.~\ref{fig:isoq30}, Speed).
The remaining difference may be due to the metallicity dependence of ISO production that we assume.
We mark the asymptotic speeds of 1I, 2I, and example Solar System comet C/1956 R1 (\S~\ref{sec:known}) on Figure~\ref{fig:isoq5}. 
Again, the two known ISOs are entirely typical relative to this distribution.
Our predicted distribution gives a \textit{p}-value of \(10^{-5}\) for an ISO having a speed relative to the Sun lower than that of the weakly hyperbolic C/1956 R1 --- making this comet's Solar System origin and Oort cloud membership significantly preferred.
This reproduces a result of \cite{Whipple_1975}, who gave an analytic expression for this probability assuming a Gaussian velocity distribution, and determined C/1956~R1 
to have a \textit{p}-value in the range \(2\times10^{-7}\) -- \(1\times10^{-4}\).

The three-dimensional velocity distribution combined with the Sun's motion generates a projected distribution of the velocities of ISOs on the sky.
The radiant is the direction of approach of an ISO on its way into the Solar system, and is equal to the angular direction of its velocity vector relative to the Sun.
This is plotted in Galactic longitude and latitude \(l\) and \(b\) in Fig.~\ref{fig:isoq5} (Radiant).
Due to the motion of the Sun relative to the LSR, the predicted radiant distribution clusters near the Solar apex, marked here by the same yellow square, as expected \citep[e.g.][]{McGlynn_1989,Stern_1990,Seligman_2018}.
Since gravitational focussing increases in an isotropic manner the rate of ISOs entering the inner Solar system that have velocities similar to that of the Sun, it has the effect of increasing the proportion of ISOs with radiants not near the apex, diluting the peak.
In this projection, the moving groups are not as easily apparent, because they are effectively in a linear clump from our point of view.

We can now consider the physical properties of the ISOs themselves.
Here our example property, after \citet{Hopkins_2023}, is ISO water mass fraction.
It should be noted that this distribution is for ISOs which formed outside the water ice line only, however as discussed in \S~\ref{sec:stars_to_isos} we expect this to be true for most ISOs.
The distribution of water mass fraction of the ISOs entering the inner Solar system within the range of the \citet{Bitsch_2020} chemical model is shown in Figure~\ref{fig:isoq5} (Composition). 
The overall distribution is broad.
For reference, the inferred water mass fraction of 2I/Borisov, the only ISO with this property somewhat constrained from production rates, is marked by a vertical green line \citep{Seligman_2022}.
We discuss composition further in \S~\ref{sec:chemodynamics}.

The age distribution of ISOs entering the inner Solar system (Figure~\ref{fig:isoq5}, Age) is very similar to that of the \textit{sine morte} stellar population.
There are a couple of effects which could cause these to differ, however neither end up having a large effect.
Firstly, stars are observed to show a correlation between their ages and velocity dispersions \citep[e.g.][]{Nordstrom_2004} and a similar correlation in ISOs would cause older, faster moving ISOs to cross the Solar system more often, be more-highly weighted by the volume sample rate (Eq.~\ref{eq:VSR}), and shift the ISO distribution towards higher ages.
This does occur and can be seen in Fig.~\ref{fig:underlyingISO} (Age), though the effect is small.
Secondly, as discussed in \S~\ref{sec:sine_morte}, the Solar neighbourhood has a flat age-metallicity relation meaning that even if ISO production correlates with metallicity as we assume, there are high-metallicity, high-ISO-production stars at all ages and the ISO and stellar age distributions will remain very similar.
ISOs of all ages are passing through the Solar System: while there is a peak at \qty{3}{\giga\year}, we predict the ISO age distribution will be sampled quite evenly from stars across the age of the Galaxy.

Finally, we also include the distribution of the metallicities of the parent stars of the ISOs entering the inner Solar system (Figure~\ref{fig:isoq5}, Metallicity).
This tends towards higher metallicities than the overall stellar distribution. 
This outcome is due to the metallicity dependence that we apply to our ISO production (\S~\ref{sec:stars_to_isos}): since in our model higher \(\MH\) stars produce more ISOs, more ISOs come from higher \(\MH\) stars.
On this distribution we have plotted the value of \(\MH\) corresponding to 2I's water mass fraction of 0.3, showing that for this physical property value, 2I would have come in our model from a star of metallicity \(\MH\approx 0.1\).

We can also now see the stellar origin of the ISOs on either side of the range of the planetesimal composition model.
Though we cannot measure the distribution of ISOs in \(\fHHO\) outside of the boundaries of the chemical model (marked by vertical turquoise lines), we can still quantify the fraction of ISOs on either side.
We show the contribution to the distribution below the range in a bin covering 0 to the lower limit (\(\fHHO=0.07\)), and above the range in a bin of the same width.
The bin above the range of the chemical model has a low density, showing the \(\fHHO\) distribution only extends a little above the range that we can measure.
On the other hand, the bin below the lower limit of the chemical model has a high density, meaning the composition distribution must have a second peak in this bin, even if it is outside of the range where we can model it.
A tiny number of ISOs come from stars with metallicities below the lower limit, whereas a more significant number come from stars with metallicities above the upper limit. 
In our model, extremely metal-poor stars contribute almost no ISOs to the Solar System neighbourhood.

\subsection{Chemodynamics}
\label{sec:chemodynamics}

\begin{figure}[t]
\centering
\includegraphics[width=\figwidth\textwidth]{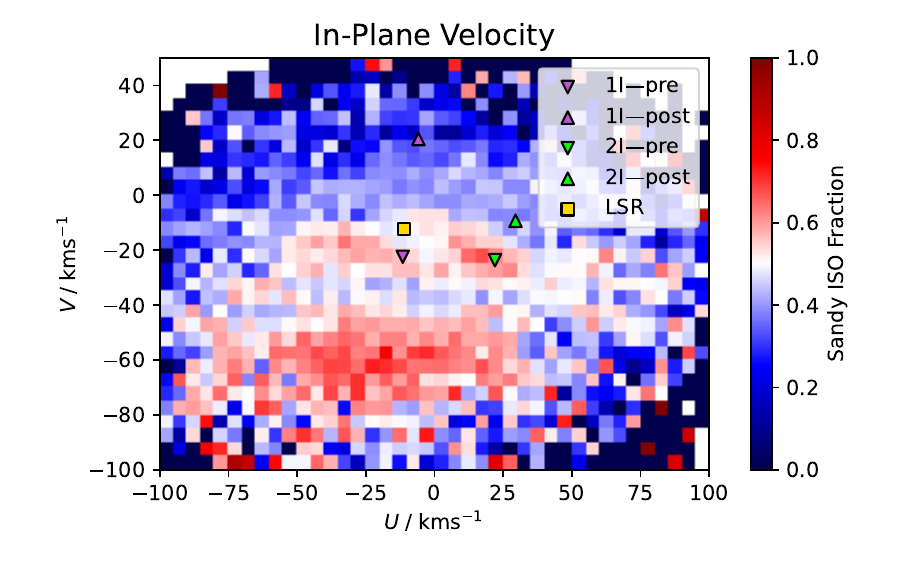}
\includegraphics[width=\figwidth\textwidth]{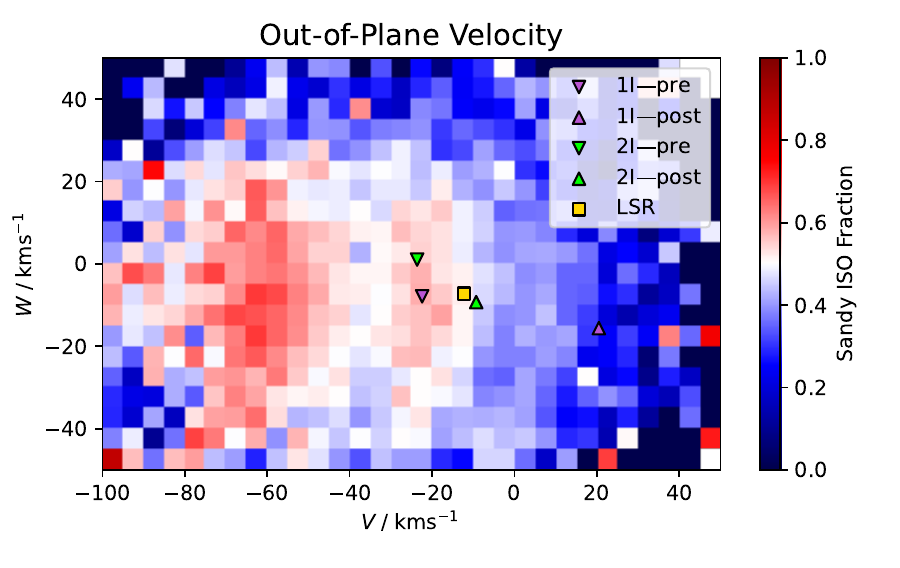}

\includegraphics[width=\figwidth\textwidth]{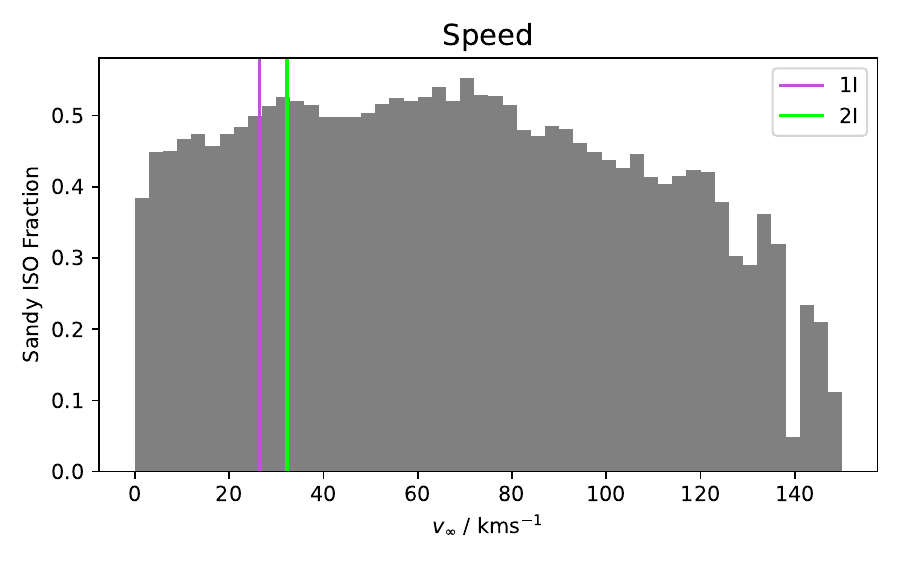}
\includegraphics[width=\figwidth\textwidth]{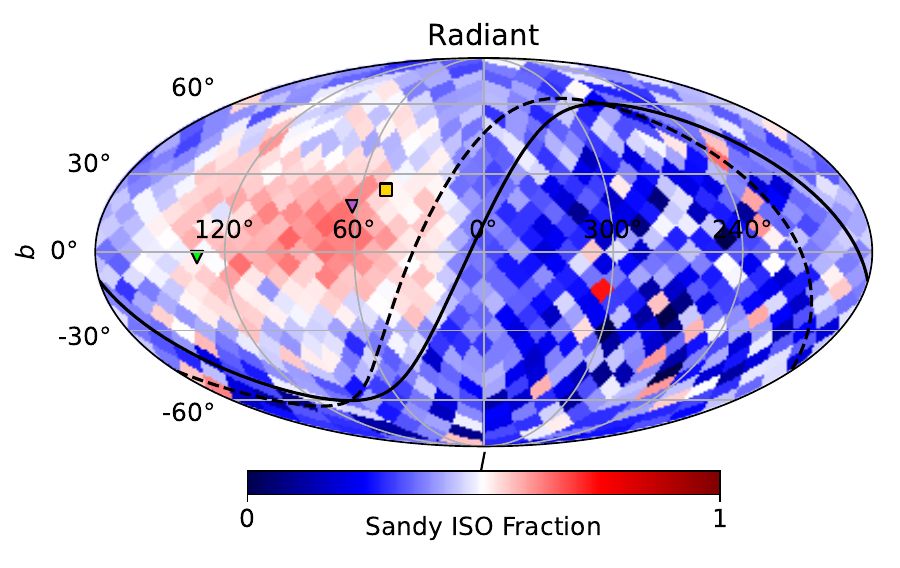}

\includegraphics[width=\figwidth\textwidth]{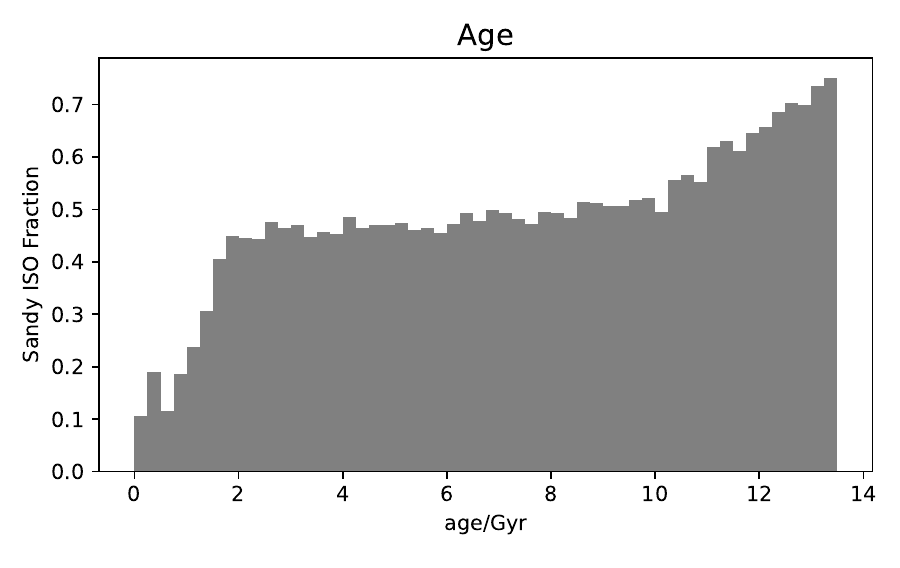}
\caption{Fraction of \(q<5\)~au ISOs at each velocity, radiant and age that are `sandy' (\(\fHHO < 0.25\)). Sandy ISOs form the majority in moving groups and low \(V\), and at higher ages.
}
\label{fig:lowf}
\end{figure}

Stars in the Milky Way exhibit correlations between composition and dynamics, so we expect ISOs with varied compositions to also have different velocity and age distributions. 
Milky Way stars are loosely grouped into a high-metallicity, low-velocity-dispersion, low-age `thin disk' and a low-metallicity, high-velocity-dispersion, high-age `thick disk' \citep{Recio-Blanco_2014}.
To explore these correlations in the ISO population, we split our predicted ISO distribution in half by water mass fraction, relative to the median \(\fHHO=0.25\).
Again this applies only to ISOs formed outside the water ice line, but we expect this to be true for most ISOs.
For ease of reading, we term these two groups `sandy' (\(\fHHO<0.25\)) and `frosty' (\(\fHHO>0.25\))\footnote{Purely compositional: no descriptor of comae grain size is implied.}. 

Since ISO water mass fraction decreases monotonically with parent star metallicity, frosty ISOs originate from \textit{low} metallicity stars, with \(\MH<0.16\), and sandy ISOs originate from \textit{high} metallicity stars, with \(\MH>0.16\).
For example, 2I would fall in the \(\fHHO>0.25\) frosty group. 
The plots in Fig.~\ref{fig:lowf} show the fraction of ISOs at each velocity, radiant and age that are sandy.
Notably sandy ISOs, like the high metallicity stars they originate from, are preferentially found in the moving groups (compare Fig.~\ref{fig:isoq5} and Fig.~\ref{fig:lowf}, In-Plane Velocity) with the sandy ISO fraction greater than 0.5 signifying that they make up the majority. 

Conversely, frosty ISOs are much more continuously distributed, and are in the majority at higher speeds relative to the Sun.
This difference is most clear in the radiant plot, which shows that the majority of ISOs entering the inner Solar system near the Solar apex will be sandy, whereas ISOs on other trajectories will be predominately frosty. 
Additionally, the age distribution changes between the two populations, with the sandy distribution being weighted towards older ISOs than the frosty population.
This is a product of the presence of old high-metallicity stars in the Solar neighbourhood (see Fig.~\ref{fig:stars}, Age-Metallicity).
Less notably, both sandy and frosty ISOs are found at all values of $v_{\infty}$, with a slight preference for frosty ISOs at higher speeds.

\section{Discussion}
\label{sec:discussion}

The correlations we show in \S~\ref{sec:results} in our chemodynamical distribution mean that substantial information about the origin of an ISO can be inferred from its velocity alone.
The conditional distribution of ISO compositions, parent-star metallicities and ages depends on the given velocity.
That is, even if we only know the velocity of an ISO, we can still calculate the probability distribution its composition, parent-star metallicity and age must have been drawn from.
These correlations will allow us to make inferences about partially characterised ISOs found in future surveys.
Our approach could also be used to constrain the origin of more exotic planetesimal ISOs found in future: if an object were not described well by our current chemical model, e.g. an icy fractal aggregate \citep{Moro-Martin_2019b}, we can still use our distribution to constrain the metallicity of its home star.

Velocity is one of the first parameters that can be measured for a new-found minor planet: for an ISO, with its comparatively high velocity relative to Solar System objects, it is quantifiable with a relatively short orbital arc, e.g. 4-7 days for LSST \citep{Cook_2016}.
Until then, orbital solutions have not yet settled, and new discoveries often sit at marginally hyperbolic ($e \sim 1.01-1.03$).
The \textit{Gaia}-based $v_\infty$ distribution we present in Fig.~\ref{fig:isoq5} can be used for testing both future ISO discoveries and known Solar System comets that have been hypothesised to have interstellar origins (e.g. C/1956 R1).
From Fig.~\ref{fig:isoq5}, the uncertainties on the velocities of discovered ISOs will need to be $\lesssim 5$~km/s for their velocity distribution identification to be crisp, as this is approximately the scale of the moving groups.

Perhaps the most striking feature of our results is that local ISOs are indeed coming from stars of all ages in the Galaxy. 
This wide age distribution is a natural outcome, given the wide stellar age distribution and the fact that ISOs outlast stars. 
This is due to radial migration; stars currently in the Solar neighbourhood have origins over a large range in Galactic radius \citep{Sellwood_2002,Lian_2022}.
Stars mix even faster azimuthally, as discussed in \S~\ref{sec:stars_to_isos}.
This means stars that happen to currently be nearby have in fact been sourced from all over the Galactic disk.

The result is that the population of stars and therefore ISOs currently near the Sun at any particular velocity is a dynamic mix of many different populations, sourced from all ages and a large swath of the disk. 
There are correlations, as the population is not fully mixed.
For instance, younger stars which have more circular orbits are more affected by resonances \citep{Daniel_2015}, and so end up in moving groups more frequently.
This however equally applies to the abundant population of low-velocity old stars. 
Unlike in purely stellar studies, ISOs have the potential for ground truth by spacecraft visits, such as Comet Interceptor \citep{Jones_2024}, together with remote telescope observation.
Our prediction of the broad age spread of ISOs would be testable with detailed isotopic ratios and compositions, from comae or  sample return missions.

\subsection{Implications for the Two Known ISOs}
\label{sec:origins}

\begin{figure}[t]
\centering
\includegraphics[width=\figwidth\textwidth]{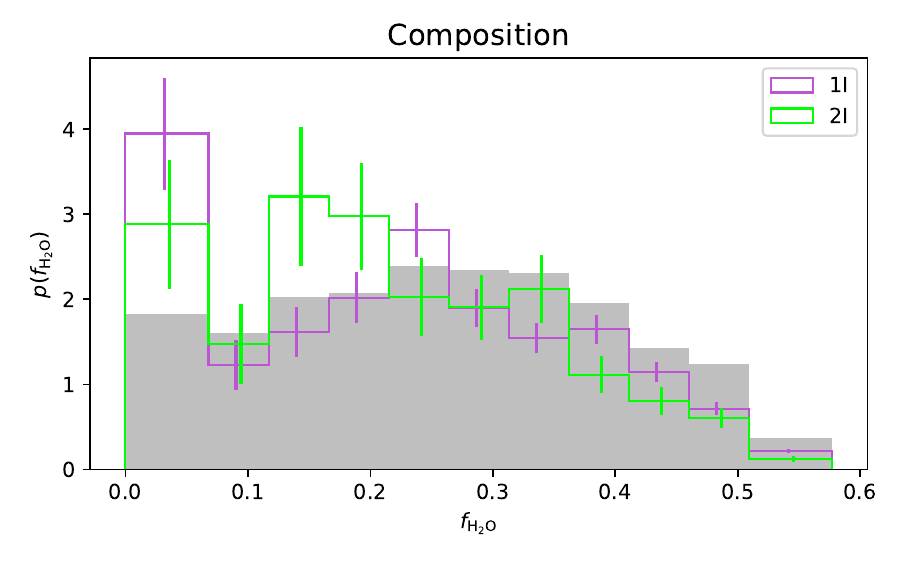}
\includegraphics[width=\figwidth\textwidth]{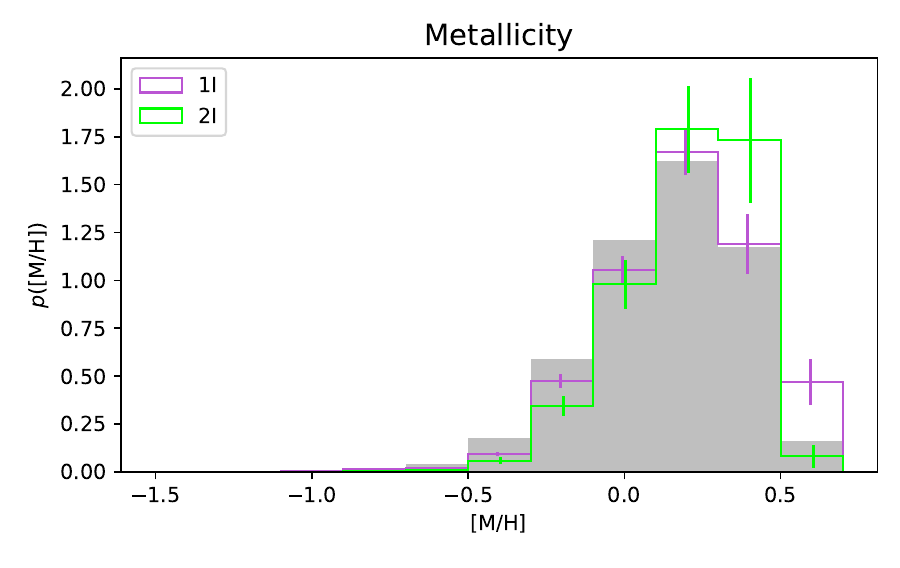}

\includegraphics[width=\figwidth\textwidth]{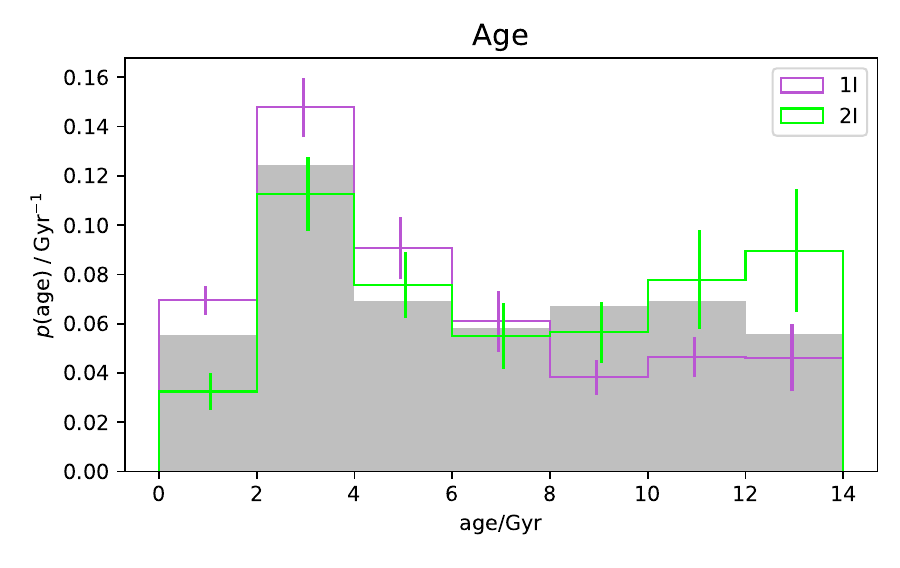}
\caption{Predicted distribution of ISOs with similar velocity to 1I and 2I in composition, age and \(\MH\). For comparison, the overall ISO population distribution in Fig.~\ref{fig:isoq5} is also shown, in grey.}
\label{fig:1I2I}
\end{figure}

The pre-encounter velocities of 1I and 2I in Fig.~\ref{fig:isoq5} both lie within the boundaries of particular overdensities in velocity space.
We reiterate that although overdensities in velocity space were originally thought to be from dispersed, coeval stellar clusters, it is now thought that they are caused by resonances with the Galaxy's transient spiral arms and bar clearing gaps in the \(U\)---\(V\) plane \citep{Skuljan_1999}.
This is supported by the \unit{\kilo\parsec}-scale extent that some of these overdensities display \citep{Ramos_2018}.

For 1I/\Ou, this overdensity is called the Pleiades moving group \citep{Zhao_2009}, part of the Hyades-Pleiades branch \citep{Skuljan_1999}, named after the Pleiades open cluster (M45) that it contains.
The Pleiades \textit{moving group} is distinct from the Pleiades \textit{open cluster}, so to avoid confusion, we use the te reo M\={a}ori name for the open cluster: Matariki \citep{Rangi_2020}.
Though \citet{Lucchini_2023} show that the Pleiades moving group has a comparatively small radial extent among moving groups, the age distribution of its members is non-coeval \citep{Antoja_2008}: this crucial point means that the moving group is far more than just the dispersal of Matariki.
Additionally, \cite{Heyl_2022} counts only \(\sim1500\) members of Matariki including escapees, but \cite{Lucchini_2023} find \(\sim\num{150000}\) stars belonging to the Pleiades moving group. 
We emphasise this distinction because 1I/\Ou's membership is of the Pleiades moving group; this does not imply formation around a star in Matariki, or at a star of one of the other young stellar associations that happen to be currently nearby \citep[e.g][]{Feng_2018} and are in the moving group.

1I/\Ou's pre-encounter velocity is also relatively near the local standard of rest, with approximately \(2\%\) of our predicted ISO distribution having a velocity closer to the LSR.
However, Fig.~\ref{fig:isoq5} (In-Plane Velocity) shows that the exact velocity of the LSR is not a preferred point in the local velocity distribution.
In particular, the structure of the branches mean there is a natural gap between the LSR and 1I/\Ou's velocity.
It is therefore not appropriate to attribute significance to the proximity of 1I to the LSR as though the velocity distribution were a Gaussian centered on the LSR.
We predict that as the sample of known ISOs increases, until 50 ISOs are known, it is most likely that 1I/\Ou\ will remain the closest to the LSR.
This is a natural outcome of the inherently clumpy distribution.

2I/Borisov was in the Coma Berenices moving group \citep[See Fig. 5 of][]{Antoja_2008} before its Solar encounter.
This moving group is named after but distinct from the constellation Coma Berenices.
Similarly, it does not represent the origin of 2I, only a transient location in velocity space caused by the clearing of the gaps by transient spiral and bar potentials.

We can use the correlations of an ISO's other properties with velocity to make inferences about the distributions that an ISO's composition, parent-star metallicity and age must be drawn from.
For ISOs discovered in the future, this means inferences can be made about their other properties based only on their velocity, which is easiest to constrain.
Here we explore what can be inferred about 1I and 2I based only on their velocities.
We do this by evaluating the composition, parent-star metallicity and age distribution of 1I and 2I using only observed \textit{Gaia} stars with velocities within \qty{5}{\kilo\meter\per\second} of each ISO's pre-encounter velocity; these are plotted in Figure~\ref{fig:1I2I}.
Setting \qty{5}{\kilo\meter\per\second} is a relatively arbitrary bound that provides a suitably large restricted sample (793 observed stars for 1I and 255 observed stars for 2I); since it is smaller than the overall sample, to ensure the differences in the histograms are significant, we calculate error bars for the value of the distribution in each bin. 
As this is a weighted sum, the variance of each value is equal to the sum of the squared weights in each bin.
Our approach uses the actual measured age and metallicity distributions of stars at the velocities of interest, but is otherwise similar to the approach used by \cite{Almeida-Fernandes_2018a}, who inferred a confidence interval on the age of 1I from its velocity using a model of the stellar age-velocity correlation.
They estimate 1I's age as \qty{200}{\mega\year}--\qty{450}{\mega\year}. 
However, their analysis uses a smooth Gaussian distribution for the stellar velocity distribution, which Fig.~\ref{fig:isoq5} shows is not representative, and their result is also dependent on the model used for the age-velocity relation.
On the validity of our own age distribution, we note that measuring ages with isochrone fitting (like \textit{Gaia}'s FLAME pipeline) can be prone to bias \citep{Nordstrom_2004}, though as noted in section \ref{sec:SF} in comparison to literature ages \cite{Fouesneau_2023} finds FLAME ages to have mean biases and dispersions of only 0.1 to 0.3 \unit{\giga\year} and \qty{0.25}{\giga\year} respectively.
Additionally, our distributions in Fig~\ref{fig:1I2I} are from samples with a large extent in space (a sphere of radius \qty{200}{\parsec}).
In future, more accurate predictions of the known ISOs' distributions could be achieved by using observed stars with both similar velocity and position.

We find that at the velocities of both 1I and 2I there are ISOs with a wide range of ages and compositions, and the distributions for both generally follow those of the overall population.
This is due to the mixing described above: at any particular velocity there are stars with a wide range of ages and metallicities.
However, the correlations with velocity cause some notable differences.
Chemically, the distributions for both ISOs are shifted towards lower \(\fHHO\) and higher parent-star metallicities.
In age, the distribution for both ISOs is broad, but 1I is shifted towards lower values, and 2I towards higher values.
This is expected from the relative speeds of 1I and 2I to the LSR given that younger stars have generally lower velocity dispersions than older, but it is important to note that stars of all ages can be found at the velocities of both ISOs.
This is backed up by previous studies showing that the Pleiades moving group is made up of stars of all Galactic ages \citep{Famaey_2008, Antoja_2008}.
Thus these results suggest that 1I is not necessarily young.

\subsection{Predictions and Model Testing by Future Surveys Such as LSST}
\label{sec:modelcomparison}
It is currently unknown how many ISOs Rubin's LSST will detect: the theoretical predictions are at a nascent state. 
Estimates have varied by orders of magnitude, spanning ones, tens and hundreds \citep[e.g.][]{Moro-Martin_2009,Cook_2016, Marceta_2023a}.
However, two points are clear: the single-image depth increase by three orders of magnitude over previous surveys \citep[e.g. Pan-STARRS,][]{Chambers_2016} will dramatically improve the detectable parameter space for ISOs; and the first two years of surveying will scan the Solar volume to initial completeness, until the rapid motion of ISOs refreshes.
Clearly, nuanced theoretical models are critical for building toward improved detection estimates.

With a major survey on the way that is expected to increase the sample size of ISOs considerably beyond the current two, we assess what number of ISOs would be needed to distinguish between models of the velocity distribution.
Note that we are not yet incorporating the actual on-sky performance of LSST or any other survey; surveys provide actual detections with a complex efficiency function. 
Also, our approach here does not provide an absolute spatial number density of ISOs; we defer this to future work.
Here we purely make a prediction of the number of ISOs needed to distinguish models.
As discussed in \S~\ref{sec:intro}, the velocity distribution of ISOs in the Solar neighbourhood has most frequently been assumed to be a smooth multivariate Gaussian.
We calculate the number of ISOs we would need in order to distinguish the realistic, featured velocity distribution we predict from a smooth Gaussian. 
As described in \cite{Hopkins_2023}, a predicted distribution of interstellar objects can be used as a likelihood function to perform inference.
The likelihood function for a Gaussian velocity distribution with a volume sampling rate included is simply the product of the Gaussian probability density function and the volume sampling rate, normalised.
We estimate the likelihood function of our \textit{Gaia} model with a \textit{k}-nearest neighbours method, based on that of \cite{Zhao_2020}: for each sample velocity we sum the normalised ISO weights of the 99 nearest neighbour points in our Gaia distribution, divided by the volume of a sphere in velocity space with radius out to the 100th nearest neighbour.
For reference, slices though the resulting likelihood function are shown in Appendix~\ref{sec:kNNcuts} (Fig.~\ref{fig:slices}).

When a survey such as Pan-STARRS or LSST finds a sample of ISOs, the likelihood of different models producing the observed velocity distribution can be calculated and compared to determine the more-likely model.
Here we calculate the number of ISOs that will be required to discriminate between different models, for each of a series of models: our Gaia model, and the Schwarzschild velocity distributions fit to M-, G- and OB-type giant stars used by \cite{Hoover_2022} and \cite{Marceta_2023b}.\footnote{The Schwarzschild distribution is a multivariate Gaussian inclined in the U-V plane by a vertex deviation \(l_v\); the parameters used by \cite{Hoover_2022} and \cite{Marceta_2023b} are listed in table 10.3 of \cite{Binney_1998}, copied from \cite{Delhaye_1965} but ultimately sourced from \citet{Parenago_1951}.}
We draw samples of different sizes \(N\) from each model velocity distribution, then calculate the log likelihood of each sample being drawn from each other model.
Repeating this with new samples, we calculate the mean and standard deviation of the log likelihoods of each pair of models.
We can then calculate the significance of each distinction for each sample size \(N\); these are all listed in Table~\ref{tab:sigmas}.
This is equivalent to a series of relative likelihood tests \citep{Akaike_1974}.
 
\begin{table}[h]
\centering
\begin{tabular}{cc|cccc|cccc} 
\tableline
Sample size & Sample model & \multicolumn{4}{c|}{Mean \(\pm\) std of log likelihood with model} & \multicolumn{4}{c}{Significance relative to sample model (\(\sigma\))} \\
&&Gaia&M&G&OB&Gaia&M&G&OB\\
\tableline
N=50 & Gaia &\(-14.0\pm0.2\)&\(-14.4\pm0.4\)&\(-14.9\pm0.7\)&\(-21.8\pm2.2\)&\(\quad\mathbf{0.0}\quad\)&\(\quad\mathbf{0.7}\quad\)&\(\quad\mathbf{1.3}\quad\)&\(\quad\mathbf{3.5}\quad\) \\ 
N=50 & M &\(-14.0\pm0.1\)&\(-13.8\pm0.2\)&\(-14.0\pm0.3\)&\(-18.5\pm1.1\)&\(0.8\)&\(0.0\)&\(0.8\)&\(4.4\) \\
N=50 & G &\(-13.6\pm0.1\)&\(-13.4\pm0.1\)&\(-13.3\pm0.2\)&\(-15.6\pm0.7\)&\(1.6\)&\(0.8\)&\(0.0\)&\(3.2\) \\
N=50 & OB &\(-12.9\pm0.1\)&\(-12.7\pm0.0\)&\(-12.2\pm0.1\)&\(-11.4\pm0.2\)&\(7.8\)&\(7.6\)&\(4.7\)&\(0.0\)\\
\tableline
N=100 & Gaia &\(-14.0\pm0.1\)&\(-14.3\pm0.3\)&\(-14.9\pm0.4\)&\(-21.5\pm1.4\)&\(\mathbf{0.0}\)&\(\mathbf{1.0}\)&\(\mathbf{1.9}\)&\(\mathbf{5.2}\) \\
N=100 & M &\(-14.0\pm0.1\)&\(-13.8\pm0.1\)&\(-14.1\pm0.2\)&\(-18.4\pm0.7\)&\(1.1\)&\(0.0\)&\(1.2\)&\(6.3\) \\
N=100 & G &\(-13.6\pm0.1\)&\(-13.4\pm0.1\)&\(-13.3\pm0.1\)&\(-15.6\pm0.5\)&\(2.3\)&\(1.1\)&\(0.0\)&\(4.5\) \\
N=100 & OB &\(-12.9\pm0.1\)&\(-12.7\pm0.0\)&\(-12.2\pm0.0\)&\(-11.4\pm0.1\)&\(12.0\)&\(11.3\)&\(7.1\)&\(0.0\) \\
\tableline
N=150 & Gaia &\(-14.0\pm0.1\)&\(-14.4\pm0.2\)&\(-14.9\pm0.4\)&\(-21.6\pm1.2\)&\(\mathbf{0.0}\)&\(\mathbf{1.2}\)&\(\mathbf{2.2}\)&\(\mathbf{6.2}\) \\
N=150 & M &\(-14.0\pm0.1\)&\(-13.8\pm0.1\)&\(-14.0\pm0.2\)&\(-18.4\pm0.6\)&\(1.3\)&\(0.0\)&\(1.5\)&\(8.1\) \\
N=150 & G &\(-13.6\pm0.1\)&\(-13.4\pm0.1\)&\(-13.2\pm0.1\)&\(-15.5\pm0.4\)&\(3.2\)&\(1.6\)&\(0.0\)&\(6.0\) \\
N=150 & OB &\(-12.9\pm0.0\)&\(-12.7\pm0.0\)&\(-12.2\pm0.0\)&\(-11.4\pm0.1\)&\(15.7\)&\(14.7\)&\(9.4\)&\(0.0\) \\
\tableline
\end{tabular}
\caption{Details of distinguishing models. Most relevant is the sigma values of our \textit{Gaia} model relative to the other models, as this is what we posit as the intrinsic population, so these are marked in bold.}
\label{tab:sigmas}
\end{table}

Positing that the true distribution of ISOs is what we have predicted here using \textit{Gaia}, we focus on the sigma values of the \textit{Gaia} model samples relative to the other models.
We find that at \(N=50\), a \(3\sigma\) discrepancy is detectable to the OB-type Schwarzschild distribution.
This is because this distribution has the lowest velocity dispersions, far lower than those of the overall stellar population, making it the most different to our \textit{Gaia} sample. 
Dropping the sample to \(N=20\) still reaches a discrepancy of significance \(2\sigma\) to the OB-type model.
At a larger but still reasonable sample size of \(N=100\), the discrepancy between our \textit{Gaia} sample and the OB-type model exceeds the \(5\sigma\) level.
The G-type model reaches \(2\sigma\) at \(N=50\). 
The discrepancy between our \textit{Gaia} sample and the M-type model is the lowest of the three Schwarzschild models because this model has the largest velocity dispersions, and is the closest match to our \textit{Gaia} distribution.
To reach a \(2\sigma\) level discrepancy between the featured \textit{Gaia} model and the M-type Schwarzschild model, we need a sample size of \(N=350\) ISOs.

This shows that a moderate-scale sample of ISOs, such as expected from LSST following debiasing, will distinguish between velocity models.
Indeed, it shows a need for a broad slate of theoretical models exploring the full range of production mechanisms for ISOs, which make testable predictions.
Our approach similarly applies to chemodynamic models. 
We tested an alternate model with no metallicity dependence for ISO production (e.g. \citet{Hopkins_2023} Sec. 4.2); while that appears physically unlikely, several thousand would be needed before a difference is detectable, as the resulting velocity distributions are very similar.

We suggest that the Solar apex should be targeted to improve the detection of inbound ISOs at large heliocentric distances.
Our results in section \S~\ref{sec:results} confirm that while ISOs that will enter the inner Solar system will approach from directions all across the sky, more ISOs that will enter the inner Solar system will approach from near the Solar apex, as demonstrated in Fig.~\ref{fig:isoq5}~(Radiant).
Such targets would be particularly suited for spacecraft missions, such as Comet Interceptor, if their $\Delta v$ and flyby solar aspect angle requirements are met
\citep{Seligman_2018,Sanchez_2021,Jones_2024}.
Fig.~\ref{fig:isoq5}~(Speed) implies a reasonable proportion of the ISO population may be accessible,  though we do not compute mission accessibility here.
The velocity structures mean ISOs from the moving groups will approach from near the Solar apex (Fig.~\ref{fig:isoq5}).
Our chemodynamic modelling shows more ISOs from the moving groups are sandy ISOs (Fig.~\ref{fig:lowf}).
However, we note that frosty ISOs still preferentially enter the inner Solar system from near the Solar apex.
Fortunately, half of the region of high ISO radiant density around the Solar apex is within LSST's \texttt{baseline\_v3.3\_10yrs} footprint\footnote{\url{https://community.lsst.org/t/baseline-v3-3-run-released/8042}}, so we predict that the velocity structures will be detectable by LSST. 
This region within the footprint of some $\sim2500$ deg$^2$ is thus suited for an LSST mini-survey. 
The chemodynamic structures should be detectable but may be more marginal, as they depend both on protoplanetary disk models and on observational characterisation subsequent to survey discovery.
The water mass fraction variation could produce a range in the level of coma, which would affect initial detectability.
While we do not model it here, if the chemodynamic correlation in water mass fraction extends into other volatiles, early detection of inbound frosty ISOs that would enter the inner Solar System could be possible in CO-sensitive wavelengths, e.g. with ALMA or NEO Surveyor.

\section{Conclusion}
\label{sec:conclusion}

We used chemodynamical measurements of nearby stars with \textit{Gaia}, accounting for the survey's selection function and stellar death, to build the first high-resolution model of the \textit{sine morte} stellar population in the Solar neighbourhood. 
We then combine this with models of how the Galactic ISO population relates to its parent stellar population, after \citet{Hopkins_2023}, and a protoplanetary disk chemical model assuming ISOs are largely sourced from outside the water ice line.
From this, we make a novel prediction of the chemodynamical distribution of the interstellar objects passing through the Solar System. 

Our \=Otautahi-Oxford model 
predicts that the ISO velocity distribution is highly structured into moving groups and branches, like the stellar velocity distribution, and that low-\(\fHHO\) ISOs reside mainly in moving groups.
This chemodynamic variation also causes a correlation with radiant, as ISOs approaching the Sun from a direction near the Solar apex have lower water mass fractions.
Targeting surveying near the Solar apex will allow us to detect ISOs on their way into the inner Solar system, providing targets for spacecraft missions such as Comet Interceptor.
The ISO distributions we predict here can be used to assess the properties of the home stellar population of interstellar objects within weeks after the ISOs' discovery.

The current observed ISOs are typical of the predicted population.
Both 1I and 2I belong to moving groups that are the transient products of resonances with the Galactic potential.
We find that the pre-encounter velocity of 1I/\Ou\ belongs to the Pleiades moving group (there is no preferred origin in the open cluster Matariki, M45), while 2I/Borisov belongs to the Coma Berenices moving group.
1I/\Ou\ is typical, rather than unusual --- the Hyades-Pleiades branch from which it came will be the source of 18\% of the $q<5$~au ISOs passing through the inner Solar System.
The correlations between velocity and other properties mean we can use the velocities of the known ISOs to predict the distributions their ages and compositions must have been drawn from.
We find that the composition distributions of both 1I and 2I are weighted towards lower values of \(\fHHO\) and higher parent-star metallicities compared towards the overall population.
The age distribution for 1I is shifted slightly towards younger values and for 2I is shifted towards older values; this is expected from the correlation of stellar velocity dispersion with age.
However, we find that both ISOs share their velocities with stars of all ages, so they could have any age. 

Finally, we find that in addition to being visually complex and featured, our prediction will be distinguishable from smooth Gaussian velocity distributions, e.g. those fit to M-, G- and OB-type giant stars used by \cite{Hoover_2022} and \cite{Marceta_2023b}.
Using relative likelihood tests we calculate that with a sample of 50 ISOs drawn from our predicted distribution, we could rule out the OB-type giant velocity distribution for ISOs to a \(3\sigma\) level. This significance rises to \(5\sigma\) with a sample of 100 ISOs. 
This method can be generalised to compare any model.

Almost all the ISOs we will see will have come to us from across the Galaxy and across cosmic time.
The upcoming ISO sample thus holds the possibility of independently testing both the universality of planetesimal formation theories such as the streaming instability, and the current understanding of Galactic star-formation history.
It will be up to the creation of theoretical models for planetesimal and ISO formation, a more detailed understanding of the nuances of ISO dynamics in the Galaxy, and the exquisite detection capabilities of the surveys of the 2020s and 2030s to bring this promise to reality.

\begin{acknowledgements}
We thank John C. Forbes, Rosemary Dorsey and Joe Masiero for helpful discussions, and Sarah Anderson for the suggestion of comparing C/1956 R1.

M.J.H. acknowledges support from the Science and Technology Facilities Council through grant ST/W507726/1.
M.T.B. appreciates support by the Rutherford Discovery Fellowships from New Zealand Government funding, administered by the Royal Society Te Ap\={a}rangi.

This work has made use of data from the European Space Agency (ESA) mission
{\it Gaia} (\url{https://www.cosmos.esa.int/gaia}), processed by the {\it Gaia}
Data Processing and Analysis Consortium (DPAC,
\url{https://www.cosmos.esa.int/web/gaia/dpac/consortium}). Funding for the DPAC
has been provided by national institutions, in particular the institutions
participating in the {\it Gaia} Multilateral Agreement.

\end{acknowledgements}

\appendix
\restartappendixnumbering

\section{Gaia Archive Queries}\label{sec:gaiaArchiveQueries}

\smallskip\noindent
\begin{minipage}{\textwidth} 
For data:
\begin{verbatim}
SELECT s.parallax AS plx, s.parallax_over_error, s.ra AS ra, s.dec AS dec,
s.pmra AS pmra, s.pmdec AS pmdec,s.radial_velocity AS rv, s.radial_velocity_error,
ap.mh_gspspec AS mh, apsupp.age_flame_spec, apsupp.flags_flame_spec
FROM gaiadr3.gaia_source AS s JOIN gaiadr3.astrophysical_parameters AS ap USING (source_id) 
JOIN gaiadr3.astrophysical_parameters_supp as apsupp USING (source_id)
WHERE s.parallax>5 AND s.parallax_over_error>10 
AND s.radial_velocity IS NOT NULL AND s.radial_velocity_error<5
AND ap.mh_gspspec IS NOT NULL AND apsupp.age_flame_spec IS NOT NULL AND apsupp.flags_flame_spec='0'
\end{verbatim}
\end{minipage}

\medskip\noindent
\begin{minipage}{\textwidth}
For selection function \(k\):
\begin{verbatim}SELECT GAIA_HEALPIX_INDEX(2,source_id) AS healpix_,
FLOOR((gs.phot_g_mean_mag - 3)/1.0) AS phot_g_mean_mag_, FLOOR((gs.g_rp - -1.0)/0.2) AS g_rp_,
COUNT(*) AS k FROM gaiadr3.gaia_source AS gs
JOIN gaiadr3.astrophysical_parameters AS ap USING (source_id)
JOIN gaiadr3.astrophysical_parameters_supp AS apsupp USING (source_id)
WHERE gs.parallax_over_error>10 AND gs.radial_velocity_error<5 AND gs.radial_velocity IS NOT NULL
AND ap.mh_gspspec IS NOT NULL AND apsupp.age_flame_spec IS NOT NULL AND apsupp.flags_flame_spec='0'
AND gs.phot_g_mean_mag > 3 AND gs.phot_g_mean_mag < 17 
AND gs.g_rp > -1.0 AND gs.g_rp < 2.0 GROUP BY healpix_, phot_g_mean_mag_, g_rp_\end{verbatim}
\end{minipage}

\medskip\noindent
\begin{minipage}{\textwidth}
For selection function \(n\):
\begin{verbatim}
SELECT sq.healpix_, sq.phot_g_mean_mag_, sq.g_rp_, COUNT(*) AS n
FROM ( SELECT GAIA_HEALPIX_INDEX(2,source_id) AS healpix_,
FLOOR((gs.phot_g_mean_mag - 3)/1.0) AS phot_g_mean_mag_, 
FLOOR((gs.g_rp - -1.0)/0.2) AS g_rp_ FROM gaiadr3.gaia_source AS gs
WHERE gs.phot_g_mean_mag > 3 AND gs.phot_g_mean_mag < 17 AND gs.g_rp > -1.0 AND gs.g_rp < 2.0
) AS sq GROUP BY sq.healpix_, sq.phot_g_mean_mag_, sq.g_rp_
\end{verbatim}
\end{minipage}

\section{The Effect of the Volume Sampling Rate and Gravitational Focussing}
\label{sec:grav_focussing_appendix}
\begin{figure}[p]
\centering
\includegraphics[width=\figwidth\textwidth]{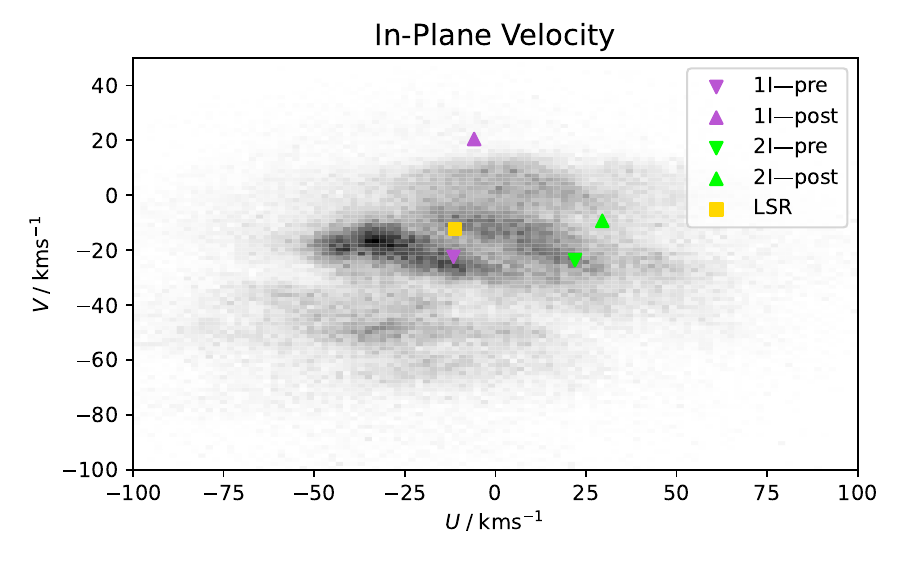}
\includegraphics[width=\figwidth\textwidth]{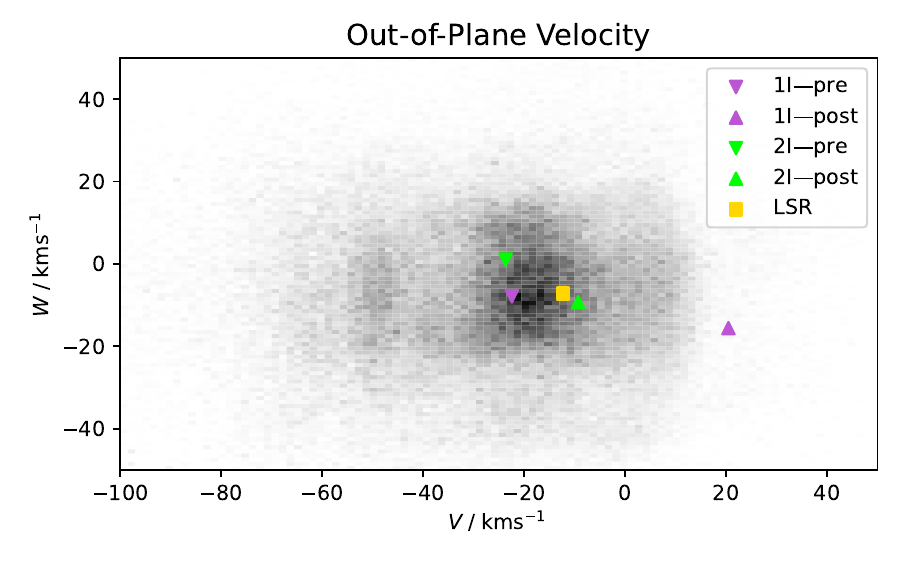}

\includegraphics[width=\figwidth\textwidth]{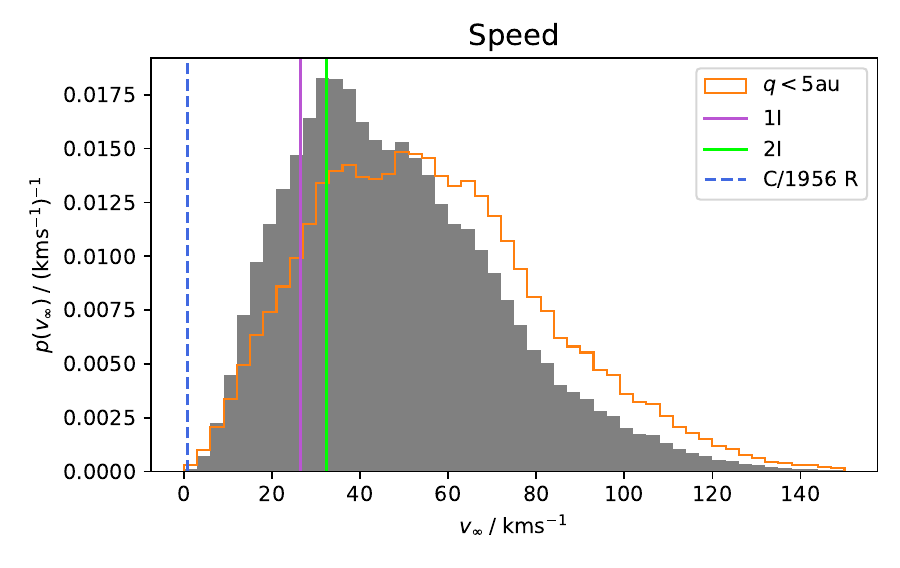}
\includegraphics[width=\figwidth\textwidth]{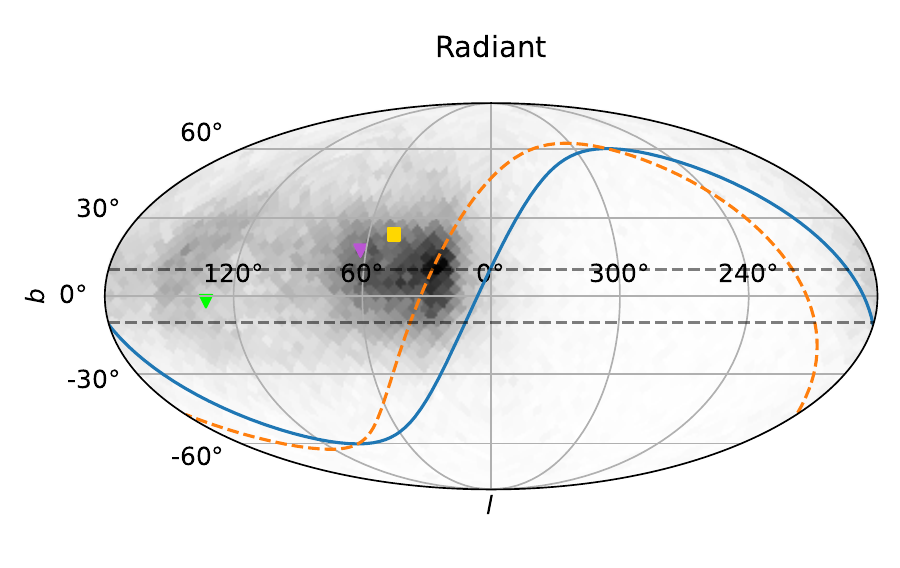}

\includegraphics[width=\figwidth\textwidth]{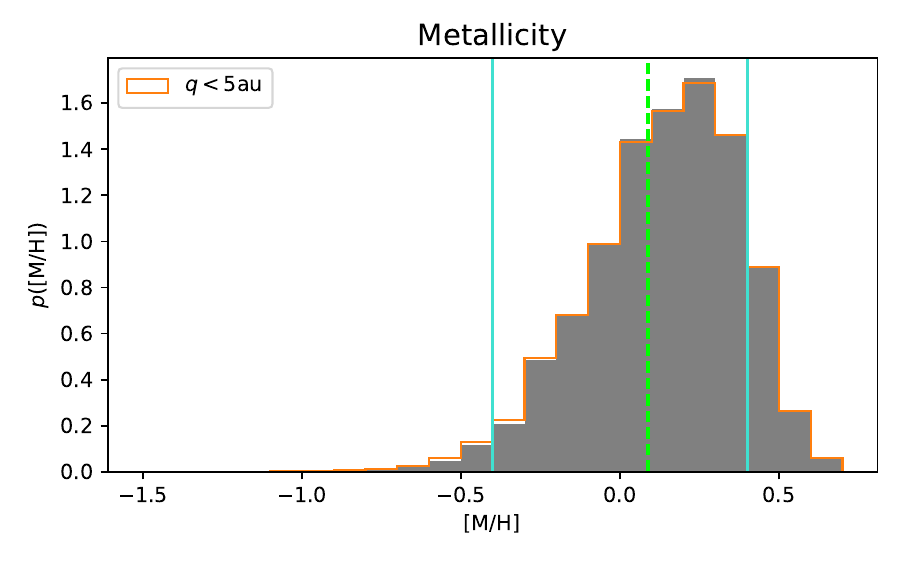}
\includegraphics[width=\figwidth\textwidth]{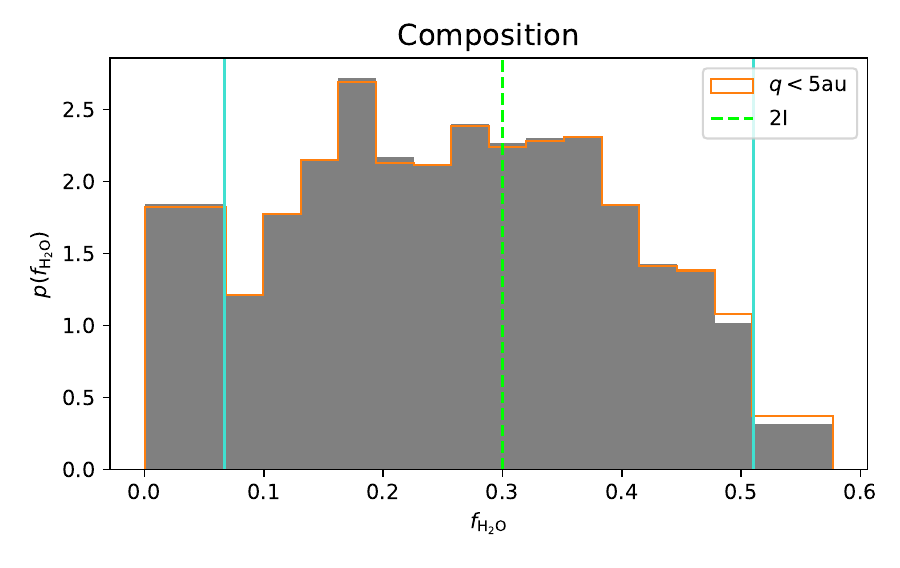}

\includegraphics[width=\figwidth\textwidth]{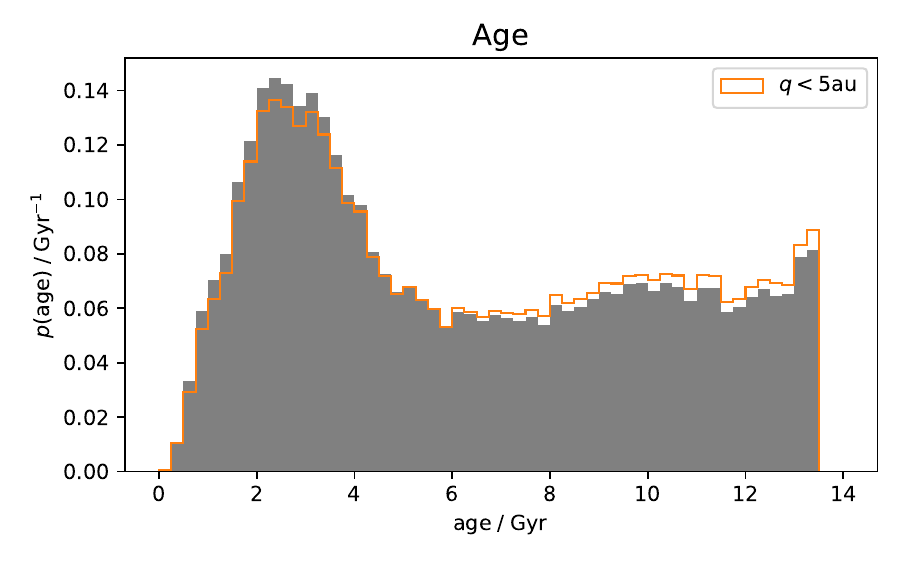}
\caption{The Solar neighbourhood ISO distribution, without including the volume sampling rate. The distribution of \(q<\qty{5}{\astronomicalunit}\) ISOs of Figure~\ref{fig:isoq5} is included as an outline for comparison---there is very little difference apart from the speed distribution.}
\label{fig:underlyingISO}
\end{figure}

\begin{figure}[p]
\centering
\includegraphics[width=\figwidth\textwidth]{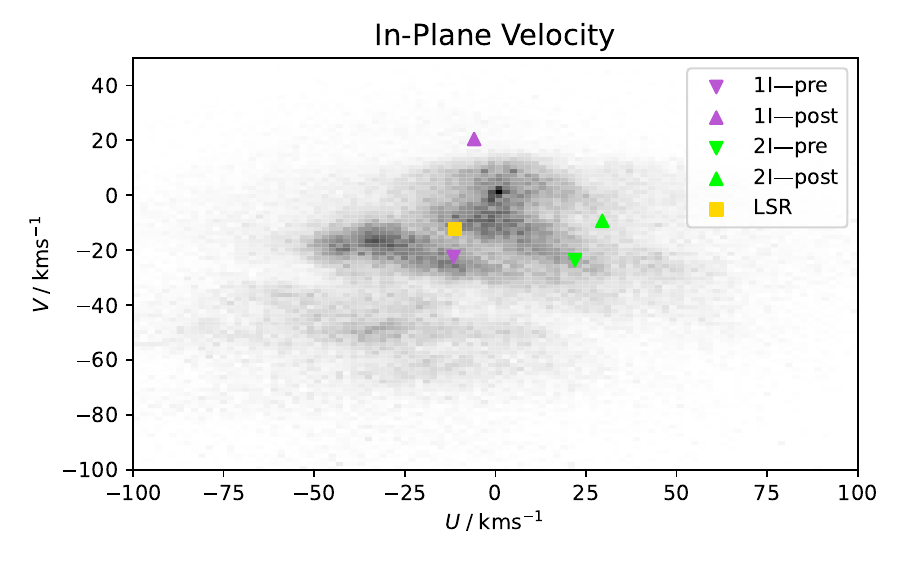}
\includegraphics[width=\figwidth\textwidth]{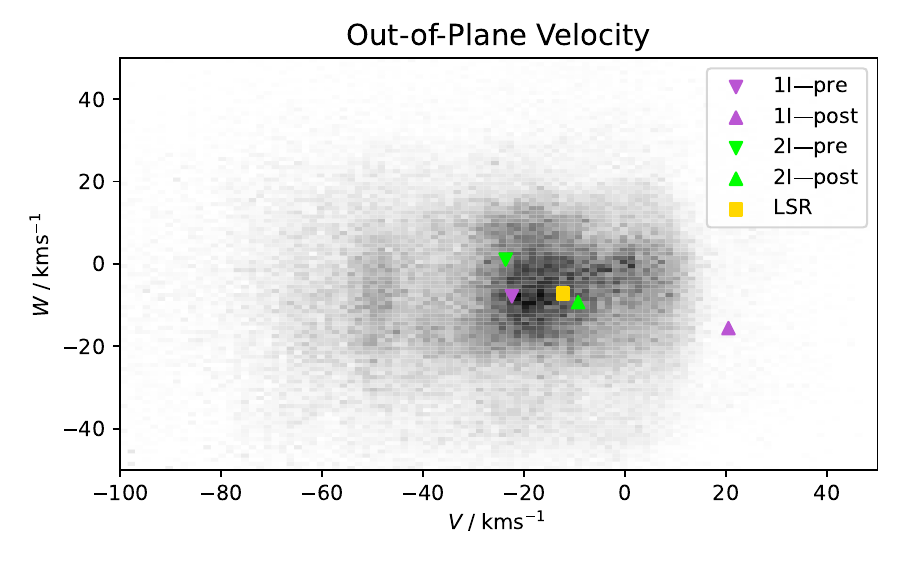}

\includegraphics[width=\figwidth\textwidth]{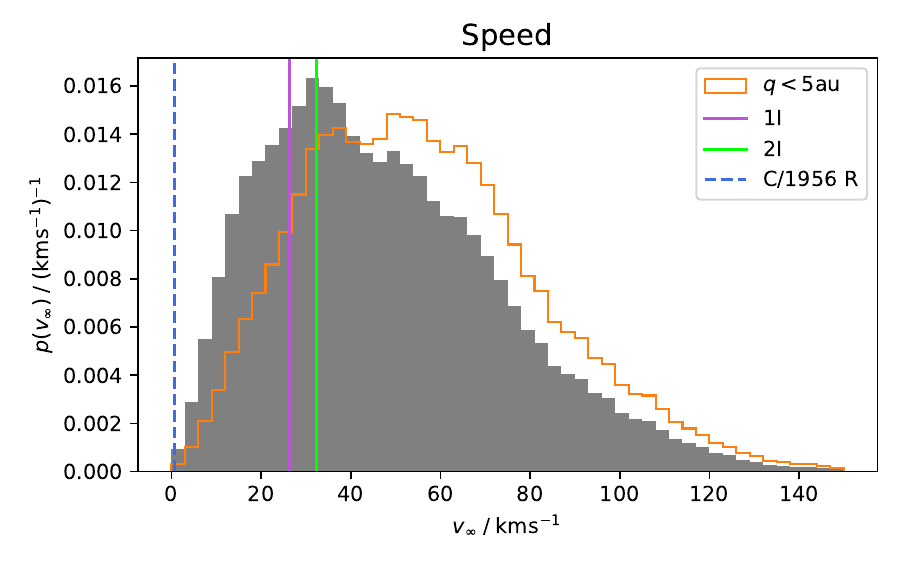}
\includegraphics[width=\figwidth\textwidth]{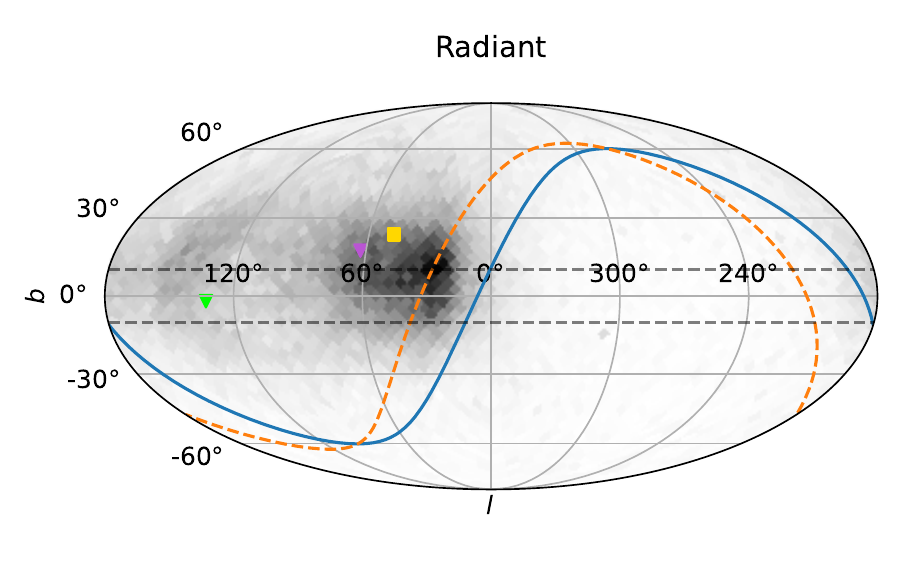}

\includegraphics[width=\figwidth\textwidth]{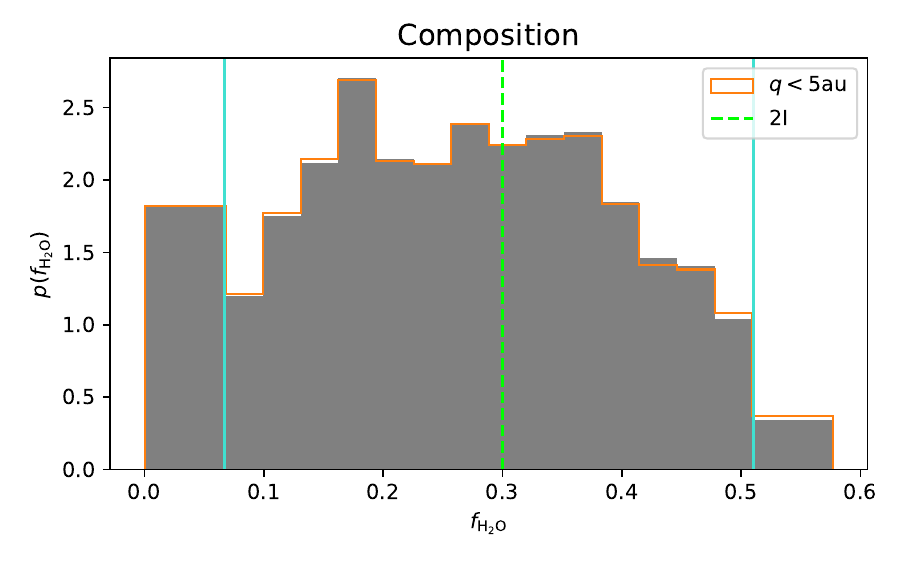}
\includegraphics[width=\figwidth\textwidth]{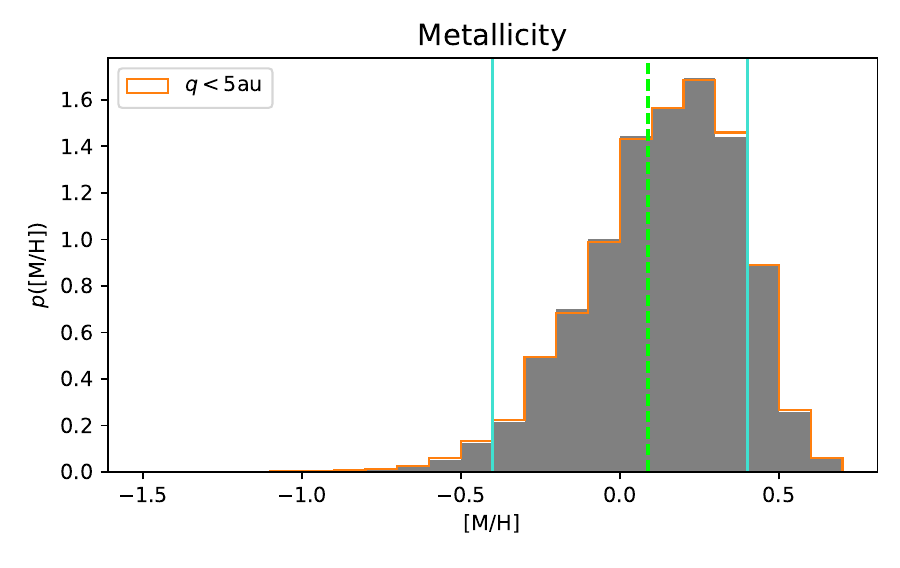}

\includegraphics[width=\figwidth\textwidth]{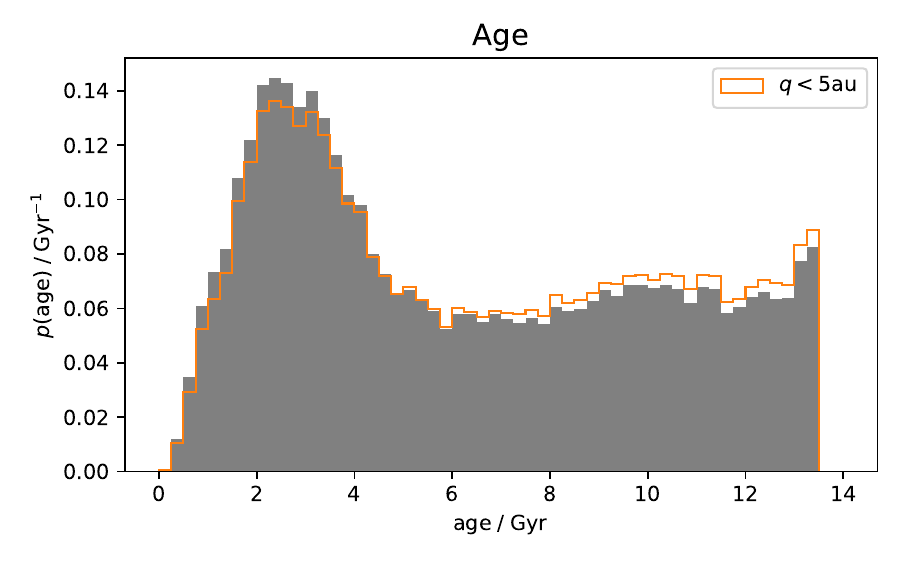}

\caption{Same as Fig.~\ref{fig:isoq5} but for ISOs with \(q<\qty{1}{\astronomicalunit}\).}
\label{fig:isoq1}
\end{figure}

\begin{figure}[p]
\centering
\includegraphics[width=\figwidth\textwidth]{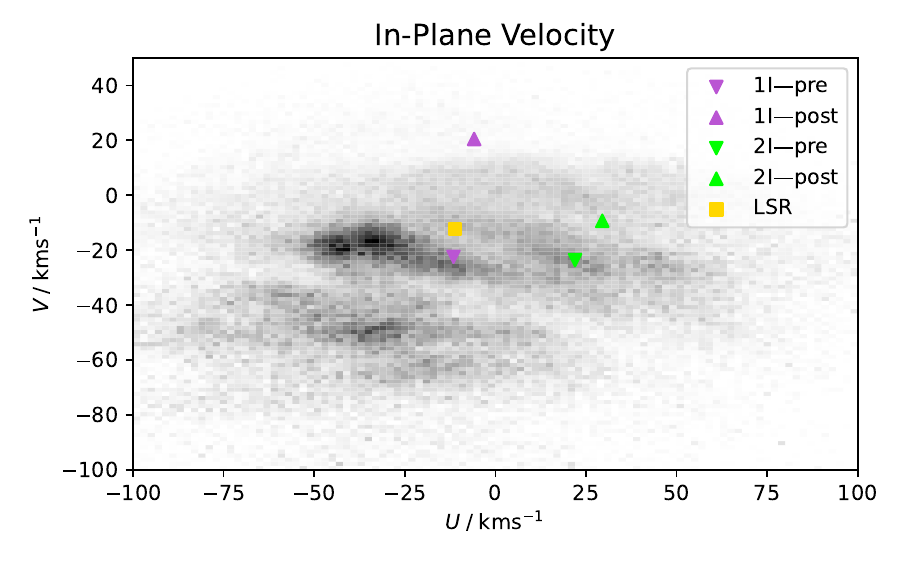}
\includegraphics[width=\figwidth\textwidth]{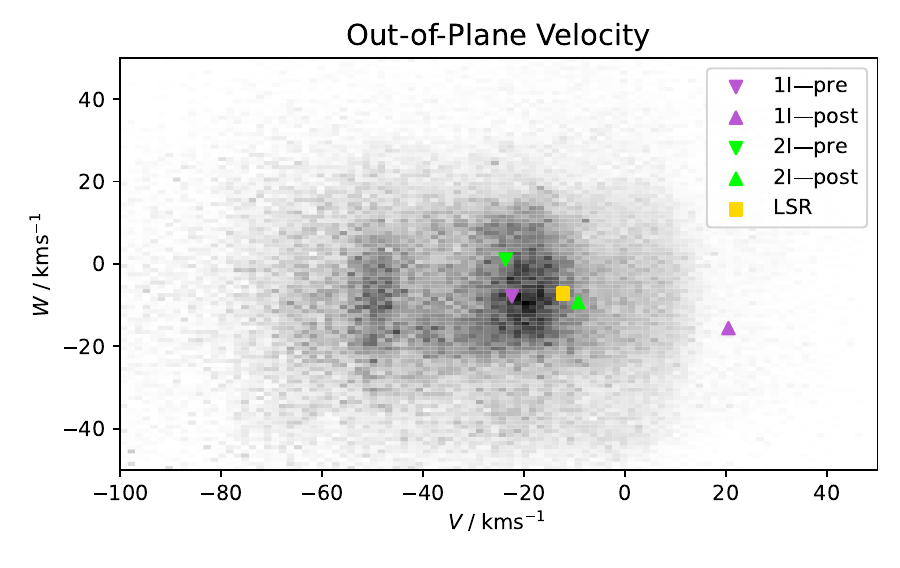}

\includegraphics[width=\figwidth\textwidth]{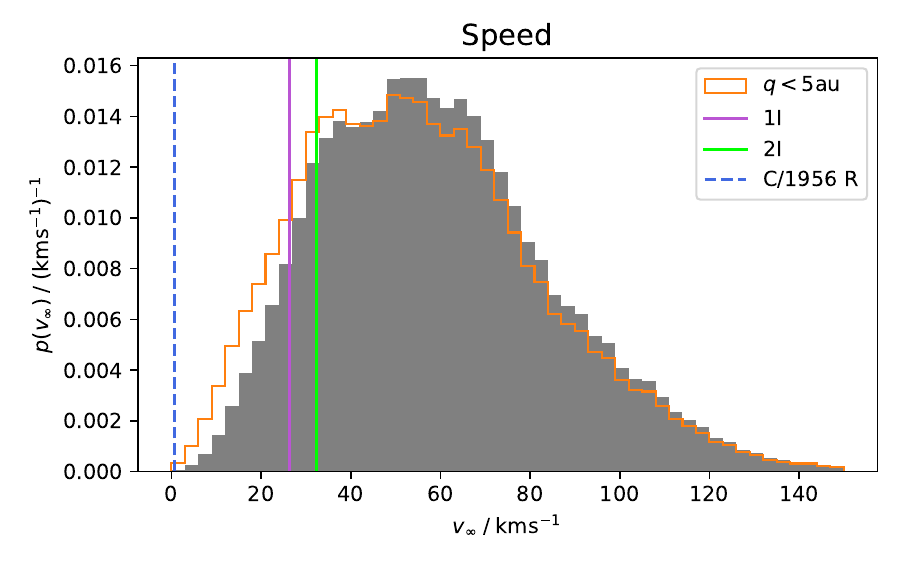}
\includegraphics[width=\figwidth\textwidth]{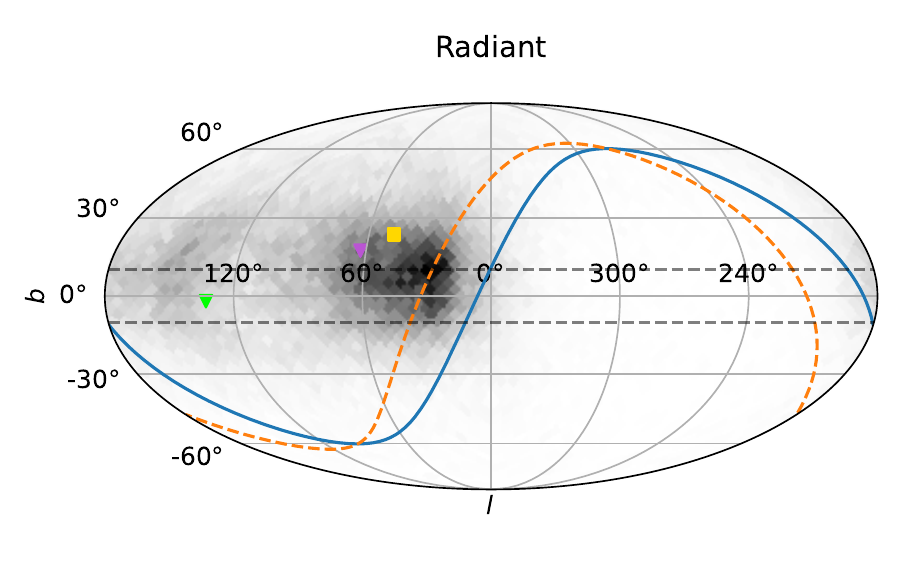}

\includegraphics[width=\figwidth\textwidth]{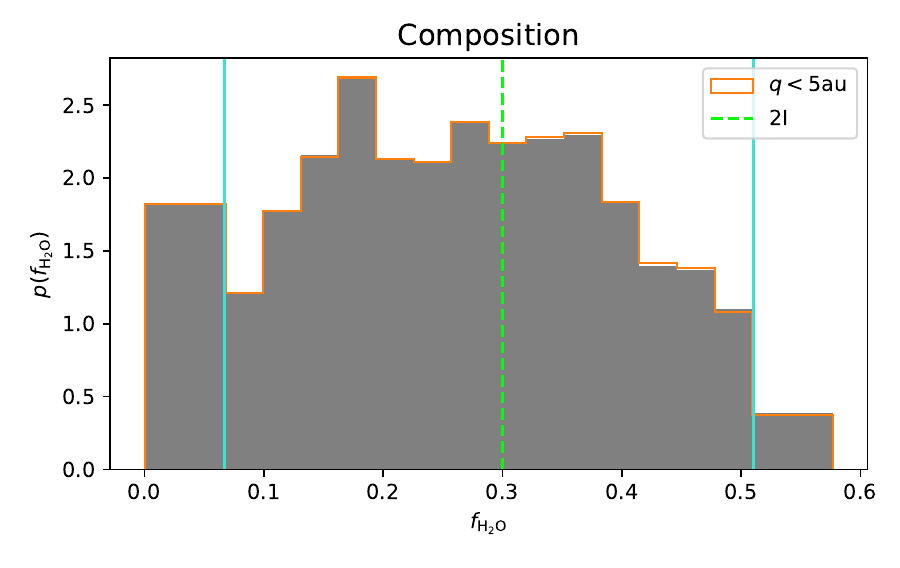}
\includegraphics[width=\figwidth\textwidth]{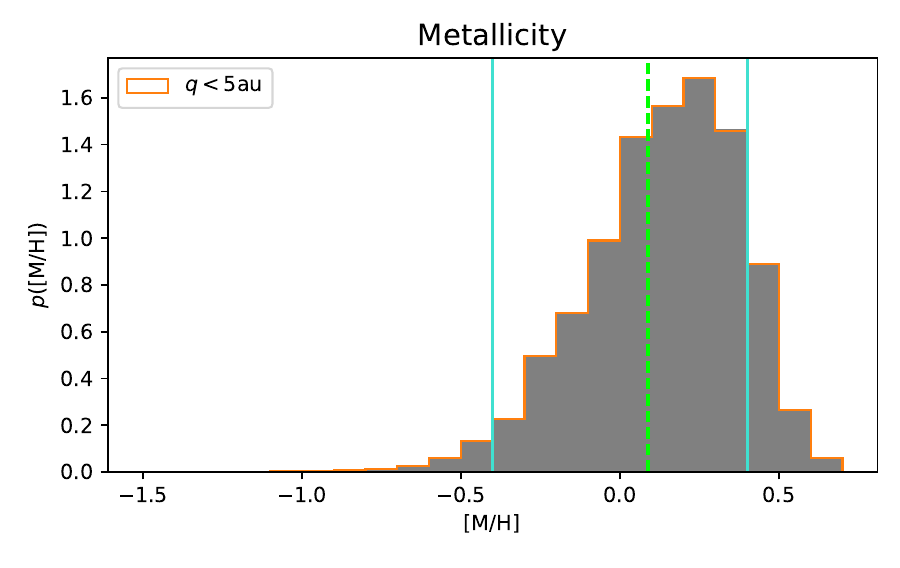}

\includegraphics[width=\figwidth\textwidth]{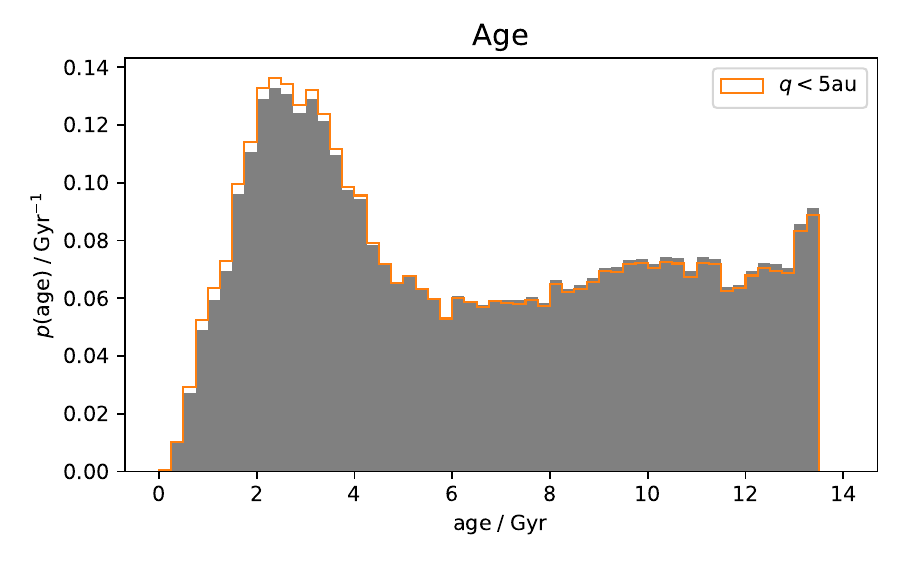}

\caption{Same as Fig.~\ref{fig:isoq5} but for ISOs with \(q<\qty{30}{\astronomicalunit}\).}
\label{fig:isoq30}
\end{figure}

In this appendix we explore how the ISO distribution depends on the volume sampling rate (\S~\ref{sec:grav_focussing}). We replot the distributions shown in Fig.~\ref{fig:isoq5} with different volume sampling rates applied: in Fig.~\ref{fig:underlyingISO} we plot the Solar neighbourhood ISO distribution, with no volume sampling rate applied. In Fig.~\ref{fig:isoq1} we apply the volume sampling rate for ISOs with perihelia less than 1~au, and in Fig.~\ref{fig:isoq30} we apply the volume sampling rate for perihelia less than 30~au.
For comparison, for each 1D histogram we also plot in an orange line our principal prediction, the ISO distributions for perihelia less than 5~au, as plotted in Fig~\ref{fig:isoq5}.

Changing the volume sampling rate changes the speed distribution as expected: in Fig.~\ref{fig:underlyingISO} the speed distribution is lowered both at high speeds (\(v_\infty>\qty{50}{\kilo\meter\per\second}\)) due to the lack of the refresh rate and at very low speeds (\(v_\infty<\qty{5}{\kilo\meter\per\second}\)) due to the lack of gravitational focussing; in Fig.~\ref{fig:isoq1} the speed distribution is shifted towards lower speeds due to the increased influence of gravitational focussing; and in Fig.~\ref{fig:isoq30} the speed distribution is shifted towards higher speeds due to the decreased influence of gravitational focussing.
There is little difference between any of the age distributions.

\section{Slices Through kNN Density Estimation}
\label{sec:kNNcuts}
\begin{figure}
\centering
\includegraphics[width=\figwidth\textwidth]{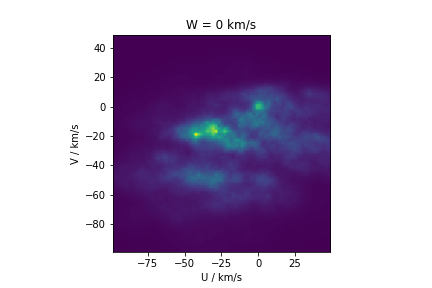}
\includegraphics[width=\figwidth\textwidth]{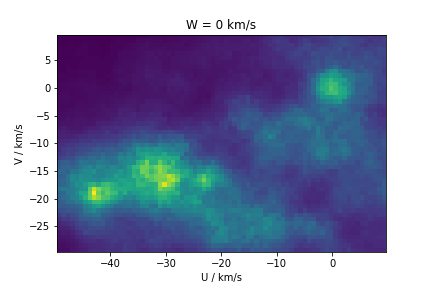}

\includegraphics[width=\figwidth\textwidth]{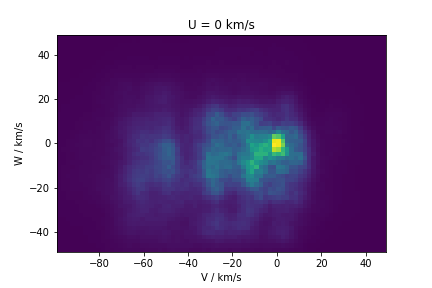}
\includegraphics[width=\figwidth\textwidth]{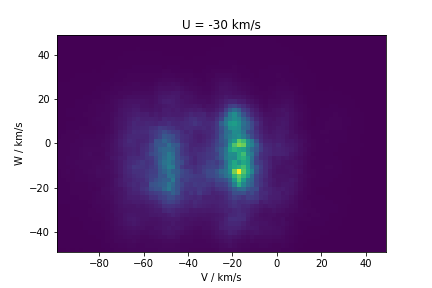}

\includegraphics[width=\figwidth\textwidth]{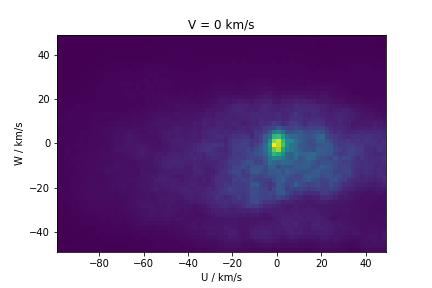}
\includegraphics[width=\figwidth\textwidth]{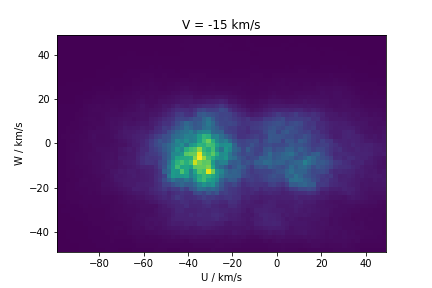}
\caption{Slices through the continuous 3D velocity distribution of \(q<\qty{5}{\astronomicalunit}\) ISOs, calculated with the kNN method described in Sec.~\ref{sec:modelcomparison}. Slices through the Sun's velocity (\(U,V,W = 0,0,0\)) show a peak caused by gravitational focussing not present in other slices.}
\label{fig:slices}
\end{figure}

This appendix comprises Figure~\ref{fig:slices}, showing slices through the kNN density estimation of the 3D velocity distribution of \(q<5\)~au ISOs used in \S~\ref{sec:modelcomparison}.

\bibliography{export-bibtex,non-ads}
\bibliographystyle{aasjournal}
\end{document}